\documentclass[11pt,letterpaper]{article}

\usepackage[T1]{fontenc}
\usepackage[utf8]{inputenc}
\usepackage{mathptmx}          
\usepackage{helvet}            
\usepackage{amsmath,amssymb}
\usepackage{booktabs,array,tabularx,graphicx,multirow}
\usepackage[table]{xcolor}
\usepackage{caption}
\usepackage{microtype}
\usepackage{geometry}
\usepackage{enumitem}
\usepackage{placeins}
\usepackage{tikz}              
\usepackage{textcomp}          
\usepackage[hidelinks,pdftitle={Consumer-Time Validity in Stateful AI Runtimes: Verifying Recovered State and Deferred Authority}]{hyperref}

\newcommand{\op}[1]{\operatorname{\mathsf{#1}}}
\newcommand{\subhead}[1]{\par\noindent\textbf{#1.}\space}
\ExplSyntaxOn
\NewDocumentCommand{\code}{m}
  {
    \group_begin:
    \fontfamily{lmtt}\selectfont
    \tl_set:Nn \l_tmpa_tl {#1}
    \tl_replace_all:Nnn \l_tmpa_tl { / } { / \allowbreak }
    \tl_replace_all:Nnn \l_tmpa_tl { _ } { \allowbreak \_ }
    \tl_set:Nn \l_tmpb_tl { }
    \tl_map_inline:Nn \l_tmpa_tl
      {
        \tl_put_right:Nn \l_tmpb_tl {##1}
        \int_compare:nNnT { \tl_count:N \l_tmpb_tl } > { 7 }
          {
            \tl_put_right:Nn \l_tmpb_tl { \allowbreak }
          }
      }
    \tl_use:N \l_tmpb_tl
    \group_end:
  }

\ExplSyntaxOff
\newcommand{\clk}[1]{\ensuremath{t_{\mathrm{#1}}}}

\usepackage{amsthm}
\theoremstyle{plain}
\newtheorem{proposition}{Proposition}
\newtheorem{lemma}{Lemma}
\theoremstyle{definition}
\newtheorem{definition}{Definition}

\allowdisplaybreaks

\setlist[itemize]{leftmargin=1.4em, itemsep=1pt, topsep=3pt, parsep=0pt}

\newif\ifPaperFiguresRelease
\PaperFiguresReleasefalse
\PaperFiguresReleasetrue

\newlength{\PaperFigureBoxWidth}

\newcommand{\PaperFigurePlaceholder}[2]{%
  \setlength{\PaperFigureBoxWidth}{#2}%
  \addtolength{\PaperFigureBoxWidth}{-0.06\PaperFigureBoxWidth}%
  \fbox{\begin{minipage}[c][1.6in][c]{\PaperFigureBoxWidth}%
    \centering\sffamily
    \textbf{Figure \csname PaperFigureNumber@#1\endcsname: artwork unavailable}\par\medskip
    \csname PaperFigureTitle@#1\endcsname\par\medskip
    \footnotesize Artwork file not present in this build.
  \end{minipage}}%
}

\newcommand{\PaperFigureArtwork}[3]{%
  \begingroup
  \IfFileExists{#2}{%
    \includegraphics[width=#3]{#2}%
  }{%
    \PaperFigurePlaceholder{#1}{#3}%
  }%
  \endgroup
}

\newcommand{\PaperFigureOne}[1]{%
  \ifPaperFiguresRelease
    \IfFileExists{fig1_failure_shapes.pdf}{%
      \includegraphics[width=#1]{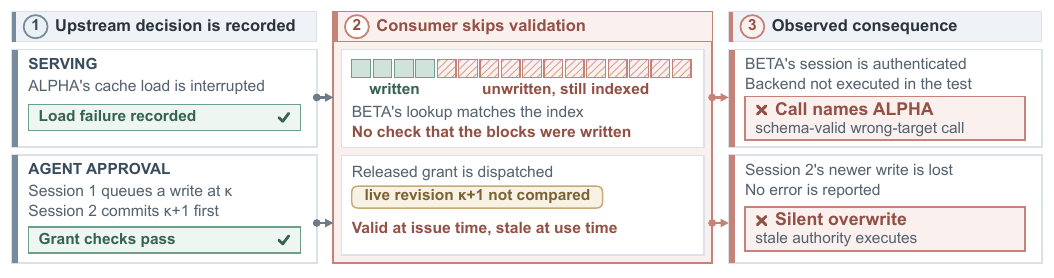}%
    }{%
      \PaperFigurePlaceholder{failure-shapes}{#1}%
    }%
  \else
    \PaperFigurePending{failure-shapes}{#1}%
  \fi
}
\newcommand{\PaperFigureTwo}[1]{%
  \ifPaperFiguresRelease
    \IfFileExists{fig2_oracle_counterexample.pdf}{%
      \includegraphics[width=#1]{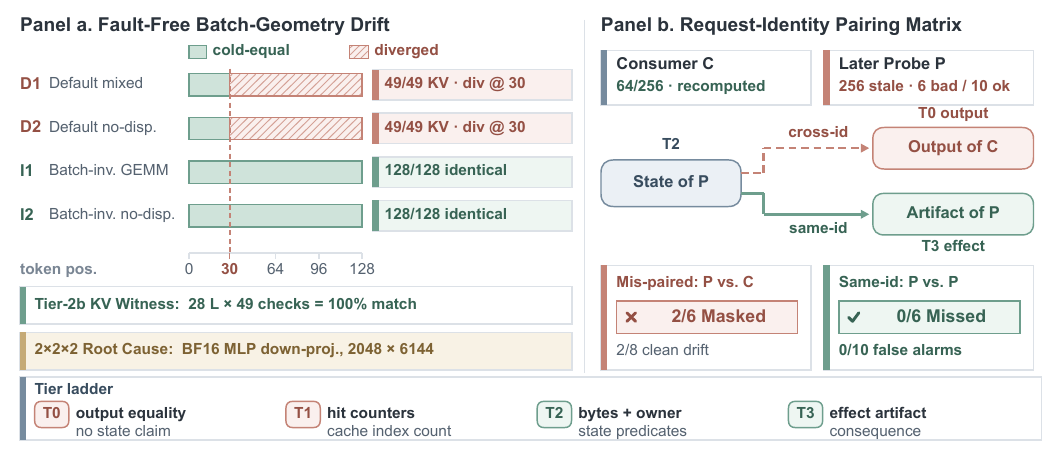}%
    }{%
      \PaperFigurePlaceholder{oracle-counterexample}{#1}%
    }%
  \else
    \PaperFigurePending{oracle-counterexample}{#1}%
  \fi
}
\newcommand{\PaperFigureThree}[1]{%
  \ifPaperFiguresRelease
    \IfFileExists{fig3_witness_architecture.pdf}{%
      \includegraphics[width=#1]{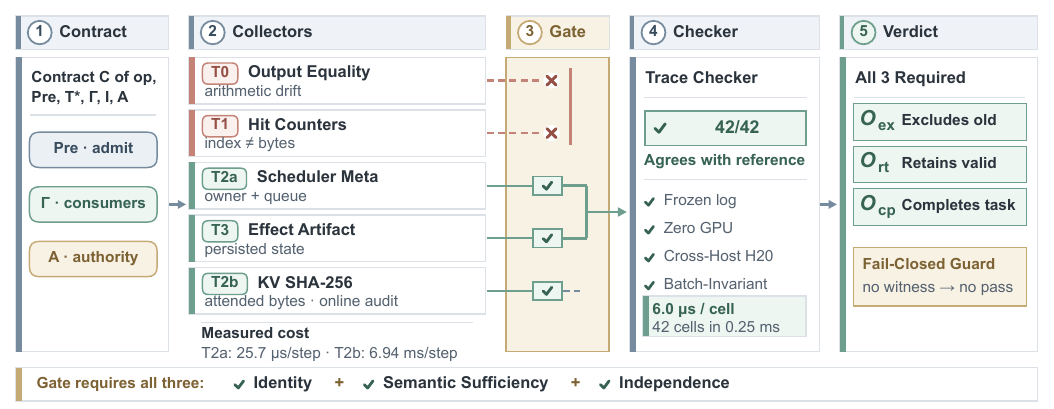}%
    }{%
      \PaperFigurePlaceholder{witness-architecture}{#1}%
    }%
  \else
    \PaperFigurePending{witness-architecture}{#1}%
  \fi
}
\newcommand{\PaperFigureFour}[1]{%
  \ifPaperFiguresRelease
    \IfFileExists{fig4_conjunctive_repair.pdf}{%
      \includegraphics[width=#1]{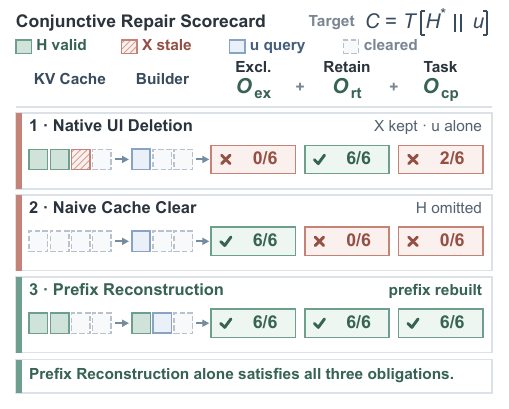}%
    }{%
      \PaperFigurePlaceholder{conjunctive-repair}{#1}%
    }%
  \else
    \PaperFigurePending{conjunctive-repair}{#1}%
  \fi
}
\newcommand{\PaperFigureFive}[1]{%
  \ifPaperFiguresRelease
    \IfFileExists{fig5_vllm_block_lifecycle.pdf}{%
      \includegraphics[width=#1]{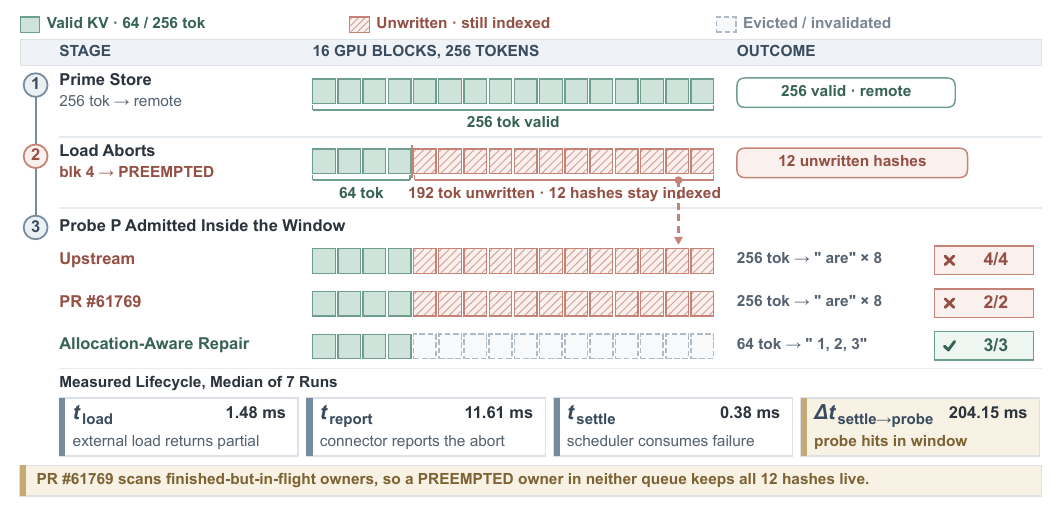}%
    }{%
      \PaperFigurePlaceholder{block-lifecycle}{#1}%
    }%
  \else
    \PaperFigurePending{block-lifecycle}{#1}%
  \fi
}

\newcommand{\PaperFigurePending}[2]{%
  \setlength{\PaperFigureBoxWidth}{#2}%
  \addtolength{\PaperFigureBoxWidth}{-0.06\PaperFigureBoxWidth}%
  \fbox{\begin{minipage}[c][\csname PaperFigureHeight@#1\endcsname][c]{\PaperFigureBoxWidth}
    \centering\sffamily
    \textbf{Figure \csname PaperFigureNumber@#1\endcsname: artwork pending}\par\medskip
    \csname PaperFigureTitle@#1\endcsname\par\medskip
    \footnotesize Internal draft placeholder.
  \end{minipage}}%
}

\newcommand{\PaperFigure}[1]{%
  \ifPaperFiguresRelease
    \csname PaperFigureDispatch@#1\endcsname
      {\csname PaperFigureWidth@#1\endcsname}%
  \else
    \PaperFigurePending{#1}{\csname PaperFigureWidth@#1\endcsname}%
  \fi
}
\newcommand{\PaperFigureAt}[2]{%
  \ifPaperFiguresRelease
    \IfFileExists{fig1_failure_shapes.pdf}{%
      \includegraphics[width=#2]{fig1_failure_shapes.pdf}%
    }{%
      \PaperFigurePlaceholder{failure-shapes}{#2}%
    }%
  \else
    \setlength{\PaperFigureBoxWidth}{#2}%
    \addtolength{\PaperFigureBoxWidth}{-0.06\PaperFigureBoxWidth}%
    \fbox{\begin{minipage}[c][1.4in][c]{\PaperFigureBoxWidth}
      \centering\sffamily
      \textbf{Figure \csname PaperFigureNumber@failure-shapes\endcsname: artwork pending}\par\medskip
      \csname PaperFigureTitle@failure-shapes\endcsname\par\medskip
      \footnotesize Internal draft placeholder.
    \end{minipage}}%
  \fi
}
\expandafter\def\csname PaperFigureDispatch@failure-shapes\endcsname{\PaperFigureOne}
\expandafter\def\csname PaperFigureWidth@failure-shapes\endcsname{\textwidth}
\expandafter\def\csname PaperFigureHeight@failure-shapes\endcsname{1.85in}
\expandafter\def\csname PaperFigureNumber@failure-shapes\endcsname{1}
\expandafter\def\csname PaperFigureTitle@failure-shapes\endcsname{Scenario, missing validation, and security consequence}
\expandafter\def\csname PaperFigureDispatch@oracle-counterexample\endcsname{\PaperFigureTwo}
\expandafter\def\csname PaperFigureWidth@oracle-counterexample\endcsname{\textwidth}
\expandafter\def\csname PaperFigureHeight@oracle-counterexample\endcsname{2.70in}
\expandafter\def\csname PaperFigureNumber@oracle-counterexample\endcsname{2}
\expandafter\def\csname PaperFigureTitle@oracle-counterexample\endcsname{The default oracle is inadmissible for state predicates}
\expandafter\def\csname PaperFigureDispatch@witness-architecture\endcsname{\PaperFigureThree}
\expandafter\def\csname PaperFigureWidth@witness-architecture\endcsname{\textwidth}
\expandafter\def\csname PaperFigureHeight@witness-architecture\endcsname{2.05in}
\expandafter\def\csname PaperFigureNumber@witness-architecture\endcsname{3}
\expandafter\def\csname PaperFigureTitle@witness-architecture\endcsname{The witness-first offline verification architecture}
\expandafter\def\csname PaperFigureDispatch@conjunctive-repair\endcsname{\PaperFigureFour}
\expandafter\def\csname PaperFigureWidth@conjunctive-repair\endcsname{3.39in}
\expandafter\def\csname PaperFigureHeight@conjunctive-repair\endcsname{2.10in}
\expandafter\def\csname PaperFigureNumber@conjunctive-repair\endcsname{4}
\expandafter\def\csname PaperFigureTitle@conjunctive-repair\endcsname{Acceptance of a repair is conjunctive}
\expandafter\def\csname PaperFigureDispatch@block-lifecycle\endcsname{\PaperFigureFive}
\expandafter\def\csname PaperFigureWidth@block-lifecycle\endcsname{\textwidth}
\expandafter\def\csname PaperFigureHeight@block-lifecycle\endcsname{2.75in}
\expandafter\def\csname PaperFigureNumber@block-lifecycle\endcsname{5}
\expandafter\def\csname PaperFigureTitle@block-lifecycle\endcsname{The physical block lifecycle of a failed external load}

\title{Consumer-Time Validity in Stateful AI Runtimes: Verifying Recovered State and Deferred Authority}
\date{}

\definecolor{SecObsidian}{HTML}{16202A}   
\definecolor{SecSlate}{HTML}{334155}      
\definecolor{SecMuted}{HTML}{64748B}      
\definecolor{SecSage}{HTML}{265C52}       
\definecolor{SecSageTint}{HTML}{DCECE7}   
\definecolor{SecBrick}{HTML}{B85450}      
\definecolor{SecBrickTint}{HTML}{FBECEB}  
\definecolor{SecCardBg}{HTML}{F8FAFC}     
\definecolor{SecFooterBg}{HTML}{EEF3F7}   
\definecolor{SecBorder}{HTML}{CBD5E1}     

\newcommand{\logohkust}{%
  \raisebox{-0.28\height}{\includegraphics[height=0.62cm]{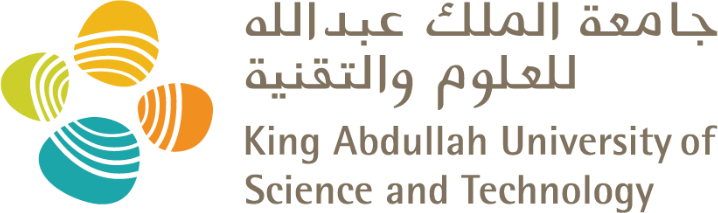}}%
}
\newcommand{\logokaust}{%
  \raisebox{-0.36\height}{\includegraphics[height=0.66cm]{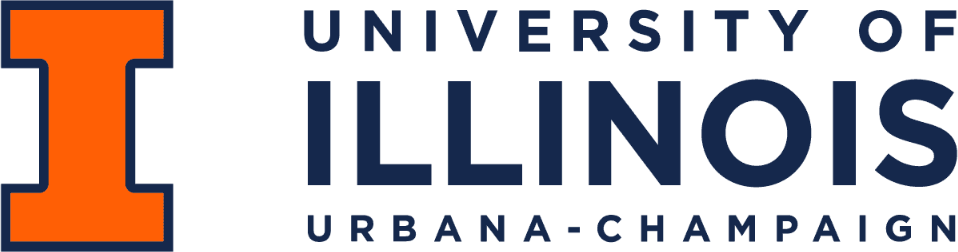}}%
}
\newcommand{\logouiuc}{%
  \raisebox{-0.36\height}{\includegraphics[height=0.66cm]{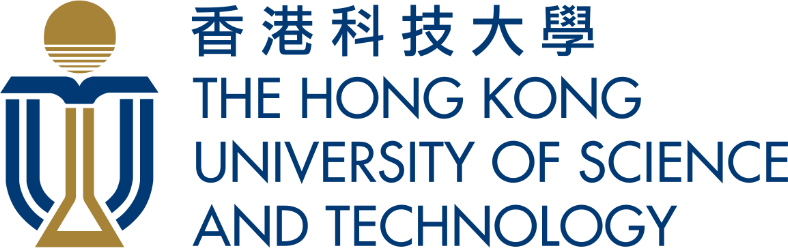}}%
}

\newif\ifuselogo
\uselogotrue

\newcommand{\dossiercard}[2]{%
  \par\noindent
  \begin{tikzpicture}
    \node[anchor=north west, inner sep=0pt, outer sep=0pt, fill=SecCardBg]
      (zone) at (0,0) {#1};
    \node[anchor=north west, inner sep=0pt, outer sep=0pt, fill=SecFooterBg]
      (foot) at (zone.south west) {#2};
    \fill[SecSage] (zone.north west) rectangle ([xshift=3.0pt]foot.south west);
    \draw[SecBorder, line width=0.6pt] (zone.north west) rectangle
                                       (foot.south east);
    \draw[SecBorder, line width=0.5pt] (zone.south west) -- (zone.south east);
  \end{tikzpicture}%
  \par
}

\long\def\blfootnote#1{%
  \begingroup
    \def\blfootnotetext{\parindent 1em\noindent}%
    \def\@currentcounter{footnote}%
    \footnotetext{\scriptsize #1}%
  \endgroup
  \hbox{}%
}

\makeatletter
\renewcommand{\@maketitle}{%
  \newpage
  \null
  \vskip 0.4em%
}
\makeatother

\begin{document}
\maketitle
\vspace{-4.2em}

%
\noindent
\begin{minipage}[b]{0.62\textwidth}
  \ifuselogo
    \makebox[\linewidth][l]{%
      \logohkust\hspace{0.9em}\logokaust\hspace{0.9em}\logouiuc}%
  \fi
\end{minipage}
\par
\vspace{0.25em}
{\color{SecObsidian}\hrule height 0.8pt}
\vspace{0.6em}

\makeatletter
\begin{center}
  {\LARGE\bfseries\color{SecObsidian} Consumer-Time Validity in Stateful AI Runtimes:\par}
  \vspace{0.15em}
  {\large\bfseries\color{SecSlate} Verifying Recovered State and Deferred Authority\par}
\end{center}
\makeatother
\vspace{0.15em}

%
%
%
\newcommand{\cormark}{\textsuperscript{\textdagger}}
\newcommand{\coremark}{\textsuperscript{\ensuremath{\bigstar}}}
\begin{center}
  {\normalsize\bfseries\color{SecObsidian}
  Guijia Zhang\textsuperscript{1}\coremark\enspace
  Huihao Jing\textsuperscript{1}\enspace
  Bing Li\textsuperscript{2}\enspace
  Shing-Chi Cheung\textsuperscript{1}\enspace
  Harry Yang\textsuperscript{1}\cormark\enspace
  Ziqi Zhang\textsuperscript{3}\cormark\par}
  \vspace{0.3em}
  {\small\color{SecSlate}
  \textsuperscript{1}The Hong Kong University of Science and Technology
  \quad
  \textsuperscript{2}King Abdullah University of Science and Technology
  \par
  \vspace{0.1em}
  \textsuperscript{3}University of Illinois Urbana-Champaign\par}
  \vspace{0.35em}
  {\scriptsize\sffamily\color{SecMuted}
  \textsuperscript{\textdagger}Corresponding author.
  \textsuperscript{\ensuremath{\bigstar}}Core contributor.\par}
\end{center}
\vspace{0.35em}

\dossiercard{%
  \begin{minipage}{\dimexpr\textwidth-12pt\relax}
    \vspace{8pt}
    \hspace{9pt}%
    \begin{minipage}{\dimexpr\textwidth-12pt-18pt\relax}
      \footnotesize
      \setlength{\baselineskip}{1.02\baselineskip}%
      \noindent
      {\scriptsize\sffamily\bfseries\color{SecSage} ABSTRACT\enspace---\enspace}%
%
%
Stateful LLM serving engines and agent runtimes reuse execution state and deferred permissions across components and sessions. A cancellation, rollback, or approval is recorded by one component and relied on by another, and recording it does not make the later use valid: a downstream consumer may read state or dispatch an action whose identity, production status, allocation, or target revision has since changed. We call the missing check \emph{consumer-time validity} and separate two forms. \emph{State validity} governs what a consumer reads. On vLLM with LMCache, an interrupted KV load leaves unwritten blocks indexed, and a host-only counter rewind desynchronizes the GPU position, so a second tenant generates a schema-valid tool call naming the wrong tenant in $10/10$ native-fault runs; the call is recorded and not executed. \emph{Authority freshness} governs what a consumer executes: on a multi-session approval runtime, an unrevoked deferred grant overwrote a newer revision in one admitted live execution. Output equality cannot test state validity: a fault-free run diverges at token 30 through BF16 batching alone while attended-KV digests match. We therefore specify \emph{consumer-bound recovery contracts} that, for state recovery, jointly require exclusion, retention, and task completion, and an \emph{admissible-witness} standard realized as a GPU-free offline trace verifier over scheduler metadata and persisted tool artifacts. On recorded traces the verifier agrees with the reference verdicts in $42/42$ frozen GPU-matrix scenarios. An allocation-aware repair satisfies $7/7$ triggered vLLM fault predicates on a pinned scheduler-case matrix, where upstream PR 61769 satisfies $2/7$, and revision-bound dispatch blocks all $160$ deterministic conflicting overwrites in a separate dispatcher cohort. Recovery must be checked where state and authority are consumed, with evidence bound to the consuming request.

      \par
      \vspace{8pt}
    \end{minipage}
    \par
  \end{minipage}%
}{%
  \begin{minipage}{\dimexpr\textwidth-12pt\relax}
    \vspace{6pt}
    \hspace{9pt}%
    \begin{minipage}{\dimexpr\textwidth-12pt-18pt\relax}
      \footnotesize
      \textbf{\color{SecObsidian}Keywords:}\enspace
      complete mediation $\cdot$ stateful AI runtimes $\cdot$ LLM serving
      $\cdot$ agent security $\cdot$ KV-cache isolation $\cdot$ TOCTOU
      $\cdot$ recovery contracts $\cdot$ verification oracles\par
      \vspace{1.5pt}
      \textbf{\color{SecObsidian}Evaluated Systems:}\enspace
      vLLM + LMCache $\cdot$ SGLang $\cdot$ Google ADK $\cdot$ ChatGLM2
      $\cdot$ DSH Approval Runtime (Cline as a control)\par
    \end{minipage}
    \par
    \vspace{6pt}
  \end{minipage}%
}
\vspace{0.65em}

\blfootnote{%
\raggedright
{\ttfamily xiongxiongchuji123@gmail.com} (G. Zhang),
{\ttfamily scc@cse.ust.hk} (S.-C. Cheung),
{\ttfamily hjingaa@connect.ust.hk} (H. Jing),
{\ttfamily yangharry@ust.hk} (H. Yang),
{\ttfamily summericeli@gmail.com} (B. Li),
{\ttfamily ziqi24@illinois.edu} (Z. Zhang).
\textsuperscript{\textdagger}Corresponding authors.}

%
\noindent
\begin{minipage}{\textwidth}
  \centering
  \PaperFigureOne{0.80\textwidth}
  {\scriptsize
  
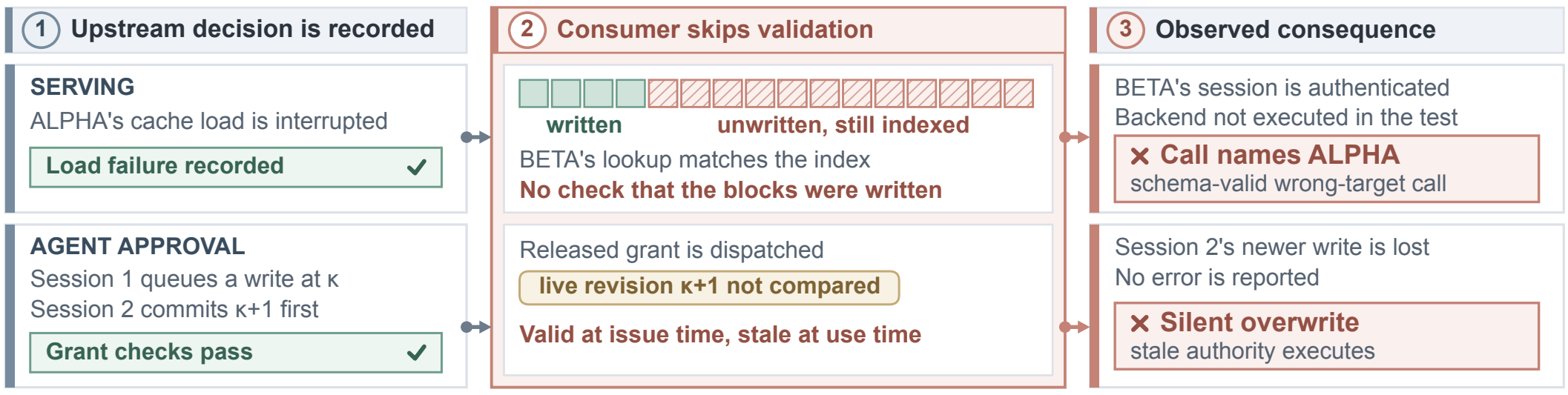
\captionof{figure}{\textbf{In both incidents the upstream decision is recorded correctly, and the consumer acts without checking it.} Top lane: ALPHA's cache load is interrupted, but the prefix index still lists blocks that were never written, so BETA's lookup matches them and BETA generates a schema-valid call naming ALPHA; no backend executes the call (Sections~\ref{sec:kv-effect} and~\ref{sec:pav}). Bottom lane: a write grant queued at revision $\kappa$ passes every identity check after Session~2 has committed $\kappa+1$, and the dispatcher never compares revisions (Section~\ref{sec:authority}). The figure is schematic. Counts and timings appear in the cited sections.}}
  \label{fig:failure-shapes}
\end{minipage}
\par

\clearpage

\section{Introduction}
\label{sec:intro}

Agent platforms no longer serve independent prompts. One conversation's execution state is spread over a transcript, a key-value (KV) cache, a prefix index shared across tenants, GPU block tables, position counters, exported memory, and queues of pending tool approvals, and platforms reuse that state to avoid recomputation~\cite{pagedattention,sglang}. Cancellation, rollback, failure recovery, and deferred approval raise a simple question: once an operation is canceled or rolled back, which state and which operations remain valid downstream? Recording the decision where it was received does not answer it for the prefix index, the asynchronous load path, the GPU position, or the tool executor that consults the state later. The answer decides whether tenants stay isolated, whether a recovered agent resumes from the right history, and whether an approved action is still authorized when it runs.

Consider tenants ALPHA and BETA that share a cached system-prompt and tool-schema prefix, loaded from an external KV store. ALPHA's load is interrupted, the connector reports the failure, and the engine follows its recovery path. A later request may safely reuse only state whose production completed and whose validity remains established. In vLLM with the LMCache connector under the \code{recompute} failure policy, this requirement is not met. The interrupted load wrote only part of the shared prefix, yet the prefix index still lists all of it. When BETA's request arrives, the index reports a full hit, and BETA is admitted onto KV positions that the aborted load never populated. In all 10 runs of our fixed fault protocol, BETA then generates a schema-valid tool call that names ALPHA as its target instead of its own tenant, although the connector failure was recorded, BETA's session was authenticated, and the call satisfied the tool schema. The call is recorded and not executed, so the experiment shows wrong-target call generation and not an unauthorized write.


The same gap appears in tool execution. In a multi-session coding agent, Session~1 queues a file write for approval at the file's revision $\kappa$, while Session~2 independently obtains approval and commits revision $\kappa+1$. Session~1's grant is never revoked, so when it is released the dispatcher finds principal, path, operation, and permission version all valid. It never reads the target's live revision, and the stale write overwrites Session~2's newer state. The approval was valid at issuance; what the dispatcher fails to re-establish is that it still applies to the file as it now stands.

In both incidents an upstream component records its decision correctly, yet the decision and its use are separated by representation, identity, and time. A token hash is not the bytes a producer wrote, and a grant names a principal and not the target's current revision. The prefix index trusts the hash, and the dispatcher trusts a grant that was valid when issued. The missing step is validation at the point of use, which we call \emph{consumer-time validity}: the state or authority that a consumer relies on must still be valid when the consumer relies on it. The property has two forms that check different things. \emph{State validity} asks whether cached or derived state may still be read after a cancellation, rewind, or block reallocation. \emph{Authority freshness} asks whether an unrevoked permission may still be exercised after its target has changed. The two share the point of use but not the identity that must be checked.

Three properties make such failures hard to anticipate, verify, and repair. \emph{The state is spread across representations and clocks}: subsystems advance the index, block tables, position counters, logs, and queues asynchronously, so a cancellation can reach one representation and miss another. \emph{The evidence that is easy to collect does not decide the question}. A fault-free run can diverge from its reference output through BF16 batching alone. A hit counter certifies an index entry and not the bytes behind it. A test of the canceled request misses the later request that consumed the state. \emph{A local repair does not give a complete recovery}: evicting an index entry, rewinding a position, discarding a cache, or terminating an owner each establishes one condition, whereas recovery must exclude invalid state, retain valid state, and let the dependent task complete.

Our method answers these properties in three steps, and each step answers a different question. A \emph{consumer-bound recovery contract} states what must hold at a consumer. A cached prefix may be read only if the producer completed it, a delayed report must still refer to the block's current allocation, and the continuation must resume from a coherent position. A queued write may run only if its target is still at the approved revision. An \emph{admissible witness} is the evidence that decides whether the contract holds for the request that consumed the state. Scheduler metadata and persisted tool artifacts qualify, whereas output equality and hit counters do not. A \emph{boundary repair} then makes the contract hold where the state is consumed, without discarding valid state. Our GPU-free offline verifier reads only admissible evidence and enforces nothing, so its verdicts depend on the instrumentation that recorded that evidence. Attended-KV digests come from a separate online audit. We use the method to test whether native recovery and the upstream patch satisfy the contract.

That recorded rollback can diverge from the state a model attends is not new. An earlier study showed that a retained session KV cache can outlive a logical rollback in application-level agents, and it placed content-addressed prefix caches such as vLLM's outside that channel~\cite{abortedbut}. Public vLLM reports describe individual serving failures and one local fix~\cite{vllmrace,vllmfix,vllmpreempt}. Our contribution begins where logical-history criteria stop: in the asynchronous serving lifecycle the tokens match and the prefix is legitimately requested, so a logical-history check passes, yet production completion, allocation identity, and position coherence can still fail at the consumer.

We evaluate pinned versions of vLLM with LMCache, SGLang, Google ADK, ChatGLM2, and the DSH approval runtime, with Cline as a negative control whose checkpoint restore behaved correctly. Each result below keeps its own denominator. The vLLM case carries the main systems argument. On a pinned scheduler matrix with seven triggered fault predicates, upstream PR 61769 satisfies 2 and our allocation-aware repair satisfies all 7, because the patch covers the direct-abort path but not preempted owners or reports aimed at a recycled block. A two-by-two factorial on the live path shows why no single repair suffices: eviction alone leaves the continuation corrupted, device rewind alone leaves unwritten state admitted, and only their conjunction meets both conditions. Without the repair, the interrupted load leads a second tenant to generate a schema-valid wrong-target call in 10 of 10 runs, and the repair restores the correct target.

The remaining results concern the evidence and the other consumers. In a fault-free control, batching alone makes the output diverge at token 30 although the attended KV digests match, so output equality cannot serve as the test. The offline verifier matches the reference's verdict on all 42 cells of the frozen GPU matrix. Both read the same recorded traces, so this is conformance and not independent ground truth. Four further cases ask the same question of other consumers. ChatGLM2 shows that clearing state is not recovery, and ADK shows that a filtered view does not bind the exporter. SGLang shows that the gap is not confined to vLLM. In DSH, an unrevoked grant overwrote newer state in one admitted live execution, and revision-bound dispatch blocked the overwrite in controlled tests.

We make three contributions.
\begin{itemize}
\item \textbf{Security characterization.} On unmodified serving and approval stacks, we show that cancellation, rollback, and approval records can be valid where issued yet unenforced where state is consumed. We separate state validity from authority freshness, trace each failure to a missing consumer-side condition, and show that the failure reaches a second tenant's generated tool call.
\item \textbf{Consumer-bound verification methodology.} We define consumer-time validity as a property of one consumer event relative to a contract, with state validity and authority freshness as two instances that check different identities. We add an admissibility standard for witnesses, which rejects output equality and hit counters for state predicates. We instantiate it as a GPU-free offline verifier over scheduler metadata and persisted receipts, with attended-KV digests left to a separate online audit.
\item \textbf{Causal evaluation of repair sufficiency.} We score an upstream patch against the contract, show with a factorial ablation that eviction and device rewind are independent and jointly required on the tested path, and show with a naive-clearing control that exclusion must not cost retention. We implement boundary repairs in five systems and derive design requirements for stateful agent runtimes (Section~\ref{sec:principles}).
\end{itemize}

\section{Threat Model and Security Goal}
\label{sec:threat-model}

\subsection{System model}
\label{sec:system}

Serving engines and agent runtimes are pipelines of components that share state across sessions, and three kinds of boundary matter. A \emph{representation boundary} separates subsystems that encode one context differently, such as GPU blocks and a prefix index, an event log and an exported memory view, or a transcript and a cache. An \emph{authority boundary} separates the policy engine that issues an approval from the dispatcher that executes it. A \emph{trust boundary} separates tenants that share infrastructure. Three common optimizations make crossing these boundaries routine: paged and disaggregated KV memory pools system prompts and tool schemas across tenants~\cite{pagedattention,sglang}, asynchronous approval queues let a target change between grant and dispatch, and checkpoint and rewind interfaces move the same context backward in time~\cite{adkissue,cline,ohlocal}.

\subsection{Trusted base, assets, and adversary}
\label{sec:tcb}
\label{sec:adversary}

We trust the GPU kernel runtime, the hardware memory allocator, the host operating system and file system under external KV tiers and session databases, and SHA-256. We assume no memory-safety corruption, kernel compromise, sandbox escape, or stolen token. Within this base, three assets need protection. \emph{Cross-tenant KV isolation} requires every attention step to read only blocks whose tokens match its prefix and whose bytes a producer fully wrote. \emph{Post-revocation context integrity} requires turns removed by deletion, rewind, or prefill rejection to stay out of later forward passes, memory exports, and tool calls, while retained history survives. \emph{Authority freshness} requires every mutating tool call to bind to the live revision $\kappa$ of its target at dispatch.

The adversary is an unprivileged session that uses only public APIs and submits benign content. It controls timing. It can abort a transfer, let a timeout expire, burst requests until the scheduler preempts a load, trigger a streaming-guardrail rollback, or hold an approval request until another session has changed the target. These mechanisms separate the recorded revocation from the later consumption of state, and the same transitions occur benignly through connector drops, cache evictions, and preemptions. Unlike work on injected prompts and poisoned memory~\cite{greshake,agentdojo,storedinjection}, our adversary supplies no malicious content. We distinguish three outcomes of increasing severity: consumption of invalid or stale state, generation of a wrong-target tool call, and actual unauthorized execution by a backend. The vLLM experiment reaches the second outcome and not the third. It uses a schema-defined tenant argument and performs no backend write. The DSH overwrite of Section~\ref{sec:authority} is an executed write, made under a grant that was stale for the target's current revision. An unauthorized write would additionally require a backend that authorizes the model-supplied tenant identifier without binding it to the authenticated principal.

Each asset is guarded by a consumer operation. The prefix-admission path and the model runner guard KV isolation, the memory exporter and later forward passes guard context integrity, and the tool dispatcher guards authority freshness. The adversary's timing control matters because it separates the upstream record from the operation that depends on it.

\subsection{Security goal}
\label{sec:scope}

Across every lifecycle transition, we require three properties. \emph{Complete Mediation}: each state read and tool dispatch verifies current validity at the consuming interface. \emph{Post-revocation isolation}: the footprint of an aborted or low-trust session stays within its own execution scope. We use the term informally, without an information-flow noninterference proof. \emph{Fail-closed isolation}: an interrupted transfer or overtaken grant defaults to refusal or clean recomputation. The three are ordered by dependence. Complete mediation is the mechanism, since a check at every consuming interface is what can keep a revoked footprint out of other scopes. Post-revocation isolation is the intended outcome of that mediation, and we test it through observable consequences and not through a proof. Fail-closed isolation fixes the default when the check cannot be completed, so a missing witness yields refusal and not admission. A missing check lets the downstream use proceed even when an upstream component has recorded the failure. The goal follows the reference-monitor tradition~\cite{saltzer1975,anderson1972,bishop1996}. Section~\ref{sec:contracts} states it per consumer, and Section~\ref{sec:related} relates it to prior mechanisms.

\section{Consumer-Bound Recovery Contracts}
\label{sec:contracts}

Complete mediation requires that every access be checked, but it does not say what the check must establish. After a cancellation, rewind, or approval, the checks that components already perform answer neighboring questions. A prefix-hash match shows shared text, not written key-value tensors. A block identifier in a failure report names a slot, not its owner after reallocation. A host-side rewind restores the scheduler's token count, not the device position. An approval records what the reviewer saw, not the revision at dispatch. A \emph{consumer} is the component that reads state or executes an action after a recovery transition: the scheduler's prefix-admission path, the model runner that continues from a position, a memory exporter, or a tool dispatcher. A consumer-bound contract states what that consumer must establish before it relies on the upstream record.

The method has three parts that answer different questions. The \emph{contract} states what must hold at a consumer. A \emph{witness} is the evidence that decides whether it holds (Section~\ref{sec:oracles}). A \emph{repair} is the change that makes it hold (Section~\ref{sec:method}). None of the three substitutes for another. A verifier that reads admissible witnesses enforces nothing, a repair that passes its tests does not show that the contract holds on other paths, and an upstream acknowledgment is none of the three.

The contract sits between two other layers (Table~\ref{tab:layers}). Security goals say what the runtime should protect. The contract turns a goal into predicates on one consumer's use of one object. Witness requirements then limit which evidence may decide those predicates. They constrain the evidence and do not define correctness.

\begin{table}[!htbp]
\centering
\caption{\textbf{Three layers connect a security goal to checkable evidence.} Repairs are not a fourth layer. They are the mechanism that makes a contract hold, and witnesses decide whether it does.}
\label{tab:layers}
\footnotesize
\setlength{\tabcolsep}{3.0pt}
\renewcommand{\arraystretch}{0.95}
\begin{tabularx}{\columnwidth}{@{}>{\raggedright\arraybackslash}p{0.19\columnwidth} >{\raggedright\arraybackslash}p{0.27\columnwidth} >{\raggedright\arraybackslash}X@{}}
\toprule
Layer & Question it answers & Content\\
\midrule
Security goals (Section~\ref{sec:scope}) & What should the runtime protect? & Complete mediation, post-revocation isolation, fail-closed isolation\\
Contract (this section) & What must hold at this consumer? & Predicates $\Phi_{\mathcal C}(e)$. For state recovery: Equations~\ref{eq:problem} and~\ref{eq:prefix}, judged by exclusion, retention, and completion\\
Witness requirements (Section~\ref{sec:oracle-admissibility}) & What evidence may decide a predicate? & Identity, Semantic Sufficiency, Independence\\
\bottomrule
\end{tabularx}
\end{table}

\subsection{What a contract specifies}
\label{sec:contract-def}

A contract for one recovery transition is the tuple
\begin{equation}
\mathcal C=(\op{op},\op{Pre},T^*,\Gamma,I,A).
\label{eq:contract}
\end{equation}
Here $\op{op}$ is the lifecycle transition, $\op{Pre}$ its precondition, $T^*$ the canonical history after recovery, $\Gamma$ the consumers it covers, $I$ the interval over which it must hold, and $A$ the authorization policy. For a deleted turn, $\op{op}$ is the deletion, $T^*$ is the transcript without that turn, and $\Gamma$ contains the model runner and any memory exporter.

Consumer-time validity is a property of one consumer event, stated relative to such a contract.

\begin{definition}[Consumer event and consumer-time validity]
\label{def:ctv}
A \emph{consumer event} is a tuple $e=(c,o,a,t)$: consumer $c$ performs operation $a$ on object $o$, and $t$ is the logical point at which $c$ validates $o$ and relies on it. The object $o$ is a state representation or an authorization grant. A contract $\mathcal C$ \emph{covers} $e$ when $c\in\Gamma$, $t\in I$, and $\op{Pre}$ holds in the execution. For a covered event, the contract fixes a finite, nonempty set $\Phi_{\mathcal C}(e)$ of predicates over $o$, its lifecycle record, and the state of its target at $t$. The event is \emph{consumer-time valid} under $\mathcal C$ when every predicate holds:
\begin{equation}
\op{CTV}_{\mathcal C}(e)\ \Longleftrightarrow\ \bigwedge_{\varphi\in\Phi_{\mathcal C}(e)}\varphi(e).
\label{eq:ctv}
\end{equation}
\end{definition}

$\op{CTV}_{\mathcal C}(e)$ is defined only for covered events, and requiring $\Phi_{\mathcal C}(e)$ to be nonempty rules out validity that holds because nothing was checked. When an operation spans an interval, $t$ marks where the check is made, and Section~\ref{sec:lifecycle} states when the predicates must keep holding afterward.

The definition fixes the form of the requirement, not its content. We instantiate a contract for one protected operation in five steps. We identify the consumer and the operation it performs, then the object it consumes, and then the lifecycle transition $\op{op}$ that can change the object's validity, together with $\op{Pre}$ and the canonical history $T^*$. From these we determine the identities the consumer must check, which the authorization policy $A$ supplies for grants. Finally we state the predicates and the evidence admissible for each (Section~\ref{sec:oracles}). Table~\ref{tab:instances} shows the result for cache admission and tool dispatch. Covering an event fixes which predicates must be shown and does not establish any of them. If no contract covers an event, or a required predicate cannot be decided from admissible evidence, consumer-time validity is not established, and the fail-closed goal of Section~\ref{sec:scope} calls for refusal or clean recomputation. We carry out this procedure for the consumers studied here. We do not claim that the predicates can be derived automatically, or completely, for arbitrary runtimes.

Equation~\ref{eq:problem} states the state-selection requirement for consumers that read state. Fast-path reuse must be indistinguishable from rebuilding the state from scratch over $T^*$. Consumers read concrete indices, block tables, and logs, so we state the requirement over the representation a consumer selects. Let $S_t$ be the runtime state at time $t$. For consumer $c\in\Gamma$ and query $u$, let $\op{Read}_c(S_t,u)$ be the representation that $c$ selects from $S_t$, and let $\mathcal R_{\mathcal C,c}(T^*,u)$ be the representations equivalent to fresh construction from $T^*$ and $u$:
\begin{equation}
\op{Pre}\ \Longrightarrow\ \op{Read}_c(S_t,u)\in
\mathcal R_{\mathcal C,c}(T^*,u),\qquad c\in\Gamma,\ t\in I.
\label{eq:problem}
\end{equation}
The equation must hold for every covered consumer and every $t\in I$ in executions where $\op{Pre}$ holds. It constrains which representation the consumer selects and nothing more. Two other equalities are easy to confuse with it. Equality of generated tokens is a property of outputs, and it can fail through arithmetic alone (Section~\ref{sec:tiering}). Equality of physical KV bytes is a property of device memory, and only the Tier-2b digests observe it.

Exclusion and retention are the two directions of the selection requirement. \emph{Exclusion} ($O_{\mathrm{ex}}$) requires that no revoked or unwritten state is selected. \emph{Retention} ($O_{\mathrm{rt}}$) requires that valid history stays selected. Correct selection does not by itself make a task succeed: a consumer can select exactly the right state, and the task can still fail on a tool error or a missing resource. We therefore judge recovery by a third, conditional obligation. \emph{Completion} ($O_{\mathrm{cp}}$) requires that tasks that depend on retained history still succeed, provided the valid resources and inputs they need remain available. It is a progress condition and not a claim of universal liveness. \emph{Recovery adequacy} is the conjunction $O_{\mathrm{ex}}\wedge O_{\mathrm{rt}}\wedge O_{\mathrm{cp}}$. Exclusion alone is trivially met by deleting everything, and a recovery that discards valid state has not recovered. We call a repair adequate only on the paths and cohorts where all three were measured. These are obligations of state recovery. They are not requirements for every authorization operation, and Section~\ref{sec:authority-contract} gives the authority-side counterpart.

\begin{proposition}[An acknowledgment does not establish consumer-time validity]
\label{prop:ack}
Let $\alpha$ be an upstream acknowledgment of $\op{op}$, such as a recorded cancellation, settlement, or approval, and let $e$ be a covered consumer event after $\op{op}$. In the pinned runtime implementations we evaluated, executions exist in which $\alpha$ holds and some predicate of $\Phi_{\mathcal C}(e)$ fails. The acknowledgment therefore does not entail $\op{CTV}_{\mathcal C}(e)$, and a consumer that admits $e$ on $\alpha$ alone can use invalid state or stale authority.
\end{proposition}

\noindent\emph{Justification.} The claim is an existence result established by exhibition on pinned stacks. A recorded connector failure coexists with indexed, unwritten positions that violate $[0,r)\subseteq V(x)$ of Equation~\ref{eq:prefix} (Section~\ref{sec:kv-effect}). An unrevoked grant satisfies Equation~\ref{eq:revocation} and violates Equation~\ref{eq:staleness} (Section~\ref{sec:authority}). A recorded rewind that is applied correctly to the filtered event view coexists with an exporter that reads the raw log and exports rewound turns in $6/9$ histories (Section~\ref{sec:pending-tail}). The proposition does not say how often this happens, and it does not extend to runtimes we did not evaluate.

\begin{proposition}[Exclusion does not stand in for recovery]
\label{prop:conjunctive}
For the state-recovery contracts evaluated here, a repair that satisfies exclusion need not satisfy retention or completion, so exclusion alone does not establish recovery adequacy. On the vLLM path, restoring production completion of the admitted prefix need not restore execution-position coherence, and the reverse also holds. No single repair or check stands in for the measured conjunction.
\end{proposition}

\noindent\emph{Justification.} The claim rests on two measured families of arms. On ChatGLM2, native deletion violates exclusion in $6/6$ conditions, naive clearing satisfies exclusion in $6/6$ and drops retention and completion to $0/6$, and prefix reconstruction satisfies all three in $6/6$ (Section~\ref{sec:consumers}). On vLLM, eviction alone bounds admission and leaves the continuation corrupted in $10/10$ runs, and device rewind alone gives a correct continuation while unwritten positions stay admitted in $10/10$ runs (Section~\ref{sec:conjunctive}). The vLLM arms separate two conditions of Equation~\ref{eq:prefix} and not the three obligations. The proposition does not say the obligations are independent in general or that they are the only ones a recovery needs, and the ChatGLM2 and vLLM paths do not exhaust the combinations of satisfied and violated conditions.

\subsection{One property, two predicates}
\label{sec:two-instances}
\label{sec:identity-spaces}

State validity and authority freshness are two instances of Definition~\ref{def:ctv}, and Table~\ref{tab:instances} sets them side by side. The consumers check identities in three spaces, each advanced by a different event. Task eligibility $\mathcal T$, advanced by cancellation, records whether the consuming task may still use an object under $T^*$. Block ownership $\mathcal B$, advanced by reallocation, holds live block instances, each tagged with its allocation generation $g$. Target authorization $\mathcal O$, advanced by workspace writes, holds the live revisions of the targets that grants name. No single epoch orders all three.

\begin{table}[!htbp]
\centering
\caption{\textbf{State validity and authority freshness differ in what the consumer must check.} Both instances ask the consumer to re-establish at use what an upstream record asserted at issue.}
\label{tab:instances}
\footnotesize
\setlength{\tabcolsep}{3.0pt}
\renewcommand{\arraystretch}{0.95}
\begin{tabularx}{\columnwidth}{@{}>{\raggedright\arraybackslash}p{0.17\columnwidth} >{\raggedright\arraybackslash}X >{\raggedright\arraybackslash}X@{}}
\toprule
 & State validity (cache admission) & Authority freshness (tool dispatch)\\
\midrule
Relies on & A cached prefix and the position it resumes from & A grant issued earlier\\
Upstream record & Load-failure report, cancel, or rewind & Approval\\
Identity checked & Task eligibility $\mathcal T$, allocation generation $g$ & Principal, session, permission version, live revision $\kappa$\\
Predicate & Equation~\ref{eq:prefix}, and a generation match for late reports & Equations~\ref{eq:revocation} and~\ref{eq:staleness}\\
Admissible evidence & Tier-2a scheduler metadata, Tier-2b bytes where collected & Grant $\kappa$ against live $\kappa$, Tier-3 artifact\\
\bottomrule
\end{tabularx}
\end{table}

Both instances share one principle. An upstream decision is treated as sufficient evidence for a downstream operation, and the conditions that the operation requires are not re-established. They differ in what those conditions are. State validity concerns an object's identity, whether its production completed, whether the execution position is coherent, and where it stands in its lifecycle. Authority freshness concerns the principal, the permission, the target, the target's revision, and whether the check and the write form one atomic step. For this reason authority freshness is not expressed through exclusion, retention, and completion. It has its own predicates (Section~\ref{sec:authority-contract}).

\subsection{State validity: prefix identity and allocation generation}
\label{sec:state-validity}

Prefix caches match requests by hashing token sequences. Equal hashes show that two prompts contain the same text and show nothing about whether a producer finished writing the corresponding key-value tensors. Fix weights, tokenizer, template, masks, positional encoding, and cache layout. Let $C=T(H^*\Vert u)$ be the canonical sequence for the history $H^*$ that $T^*$ retains and for query $u$, where $T$ applies the fixed template and tokenizer. Split $C$ into a cached prefix $C_{:r}$ of length $r$ and a continuation $C_{r:}$. Let $Q_{:r}$ be the tokens that the admitted cache entries represent. Each artifact $x$ has an owner, an allocation generation $g$, a producing execution, and a verified completed range $V(x)$. The continuation also runs at a position. The scheduler keeps a host-side position $\pi_h$ for the request, and the model runner reads a device-side position $\pi_d$. For the consumer event $e$ in which the runner continues the request after an admission of $r$ positions, let $\op{SeamOK}(e,r)$ hold when $\pi_h=\pi_d$ at the seam, the point where the continuation begins. We evaluate three conditions for cached-prefix admission and subsequent continuation: token-prefix identity, recorded host/device position agreement at the continuation boundary, and production completion over the admitted positions. An admission and its continuation satisfy these three evaluated conditions when
\begin{equation}
Q_{:r}=C_{:r},\qquad \op{SeamOK}(e,r),\qquad [0,r)\subseteq V(x).
\label{eq:prefix}
\end{equation}
Equation~\ref{eq:prefix} states three conditions on one admission and the continuation that follows it. It records the conditions this paper evaluates, and it is not a sufficient criterion for execution correctness: the three conditions do not exhaust the possible runtime lifecycle failures, and the allocation-generation match for delayed reports remains a separate requirement (below). \emph{Token identity}, $Q_{:r}=C_{:r}$, is all that a prefix hash certifies. \emph{Production completion}, $[0,r)\subseteq V(x)$, says that the producer wrote the bytes behind the matched prefix. \emph{Execution-position coherence}, $\op{SeamOK}(e,r)$, says that the continuation runs from a position on which host and device agree. It is a condition on positions and not on tokens. A recompute that rewinds the host counter and leaves the device buffer in place makes the continuation run 160 positions ahead of where the host believes it is (host 128, device 288 in the evaluated runs). Equal tokens or equal outputs do not establish it. Tokens say nothing about positions, and outputs also vary with arithmetic (Section~\ref{sec:tiering}).

We state $\op{SeamOK}$ as agreement between two recorded counters because that is what the evaluation observes. The paired Tier-2a counters flag all 8 split runs of the 14-run device cohort and none of its 6 coherent runs (Table~\ref{tab:tiering}). Agreement of two counters does not show that both equal the position that $C$ requires, and it says nothing about device memory.

An aborted transfer satisfies token identity in the index and violates production completion over its unwritten blocks, while a host-only rewind violates $\op{SeamOK}$ even when every admitted block was written. The two conditions fail independently, which Section~\ref{sec:conjunctive} tests with a factorial repair. Section~\ref{sec:tiering} scores signals for production completion against an Exclusion reference, and signals for position coherence against a Coherence reference that asks whether the continuation carries the consumer's own value. Production completion describes the admitted state and is distinct from the task completion of $O_{\mathrm{cp}}$.

Block recycling adds a hazard that Equation~\ref{eq:prefix} does not address: which block instance a late failure report may act on. A delayed report may arrive after the scheduler has freed block $b$ and reallocated it to an unrelated request. Matching the report on the bare block identifier then fails or rewinds an innocent owner. Following the lease-generation discipline of Chubby~\cite{chubby}, a report $f$ issued at generation $g$ may change live state on block instance $b^{g'}$ only if $g'=g$. This identity condition governs settlement, not admission, and the two conditions depend on each other. On the vLLM paths of Section~\ref{sec:producers}, resolving a report to the allocation that produced it lets settlement find and evict the failed owner's unwritten blocks even after that owner is preempted, and the generation match keeps a late report off a recycled block. Equation~\ref{eq:prefix} then decides whether the next admission satisfies the three evaluated conditions. If settlement misses the owner, the index still lists unwritten blocks when the next request arrives, and that admission violates production completion.

\subsection{Lifecycle phases: when consumer time falls}
\label{sec:lifecycle}

A single \code{canceled} flag cannot say when a use is valid, and a single instant does not fit every consumer. Five phases recur in the cases we study. In \emph{publication}, an upstream component records state, a completion report, or an approval. In \emph{acquisition}, the consumer obtains the object or prepares to use it, for example through a prefix lookup or the release of a queued grant. In \emph{consumption}, it reads or relies on the object. \emph{Linearization} is the point at which the operation takes effect, such as the commit of a write. \emph{Settlement and retirement} close the lifecycle: outstanding reports resolve, and old resources become safe to reuse.

The point $t$ of Definition~\ref{def:ctv} is where the check is made, and some operations also need the predicates to keep holding afterward. An asynchronous KV read needs the admitted blocks to stay valid from acquisition to the end of the read, either because the predicate holds over that interval or because a lifetime mechanism, such as quarantine until readers drain, prevents reuse. A deferred write needs the check at its linearization point, so the revision comparison belongs inside an atomic check-and-write, and a check made earlier leaves a window. A reallocated block needs two separate conditions: a generation match, so that a late report cannot act on the new allocation, and safe retirement, so that no outstanding reader still uses the old one. The experiments measure the admission-time and continuation-start conditions of Equation~\ref{eq:prefix}, the generation match on late reports, and the revision comparison at dispatch. They do not measure validity over a full asynchronous read or the draining of outstanding readers, which remain design implications (Table~\ref{tab:principles}). We use the phases as vocabulary for placing predicates, not as a lifecycle protocol.

Teardown itself spans three instants: logical revocation \clk{cancel}, settlement of asynchronous reports \clk{settle}, and physical retirement \clk{retire}. Between \clk{cancel} and \clk{settle}, publication of new cache entries stops while owner metadata stays addressable, so that late reports resolve. Between \clk{settle} and \clk{retire}, blocks stay quarantined until in-flight readers drain, which follows the grace-period idea of read-copy-update~\cite{rcu}. One flag for all three opens the preemption and reuse windows of Section~\ref{sec:producers}.

\subsection{Authority freshness: grants and revisions}
\label{sec:authority-contract}

Cache admission and failure settlement check task eligibility $\mathcal{T}$ and allocation generation $g\in\mathcal{B}$. Tool dispatch checks eligibility and the live revision $\kappa\in\mathcal{O}$ of the target through the grant $\theta=(p,s,q,v,j,\rho,h(a),\kappa)$, which binds principal, session, task generation, permission version, operation type, resource, the digest $h(a)$ of the arguments of the operation $a$ of Definition~\ref{def:ctv}, and revision. Let $\prec$ be the order in which the runtime records events. $\op{TryAdmit}(\theta)$ is the dispatcher's attempt to execute $\theta$, $\op{NewAccept}(\theta)$ is its acceptance of $\theta$ for execution, $\op{Revoke}(p,s,v)$ withdraws permission version $v$ for principal $p$ in session $s$, and $\op{Stale}(\theta,\kappa)$ is the event at which the live revision of $\rho$ moves past the revision $\kappa$ recorded in $\theta$. Two invariants govern $\theta$. Withdrawing a permission version bars later admission:
\begin{equation}
\op{Revoke}(p,s,v)\prec\op{TryAdmit}(\theta)
\Longrightarrow\neg\op{NewAccept}(\theta).
\label{eq:revocation}
\end{equation}
A grant that was never revoked still ages when another session advances resource $\rho$ past $\kappa$. Dispatch must therefore validate the revision, as optimistic concurrency control validates a read set at commit~\cite{occ}:
\begin{equation}
\op{Stale}(\theta,\kappa)\prec\op{TryAdmit}(\theta)
\Longrightarrow\neg\op{NewAccept}(\theta).
\label{eq:staleness}
\end{equation}
Revocation and revision freshness are different conditions. The approval runtime we study enforces Equation~\ref{eq:revocation}, and a dispatcher that enforces it alone passes every cancellation test and still fails every concurrent-revision test in Section~\ref{sec:authority}. Performing the check is also not enough. A comparison made before the write can still miss a commit that lands in between, so Equation~\ref{eq:staleness} protects the write only when the check and the write form one atomic step.

Authority freshness has its own counterpart of the requirement that a recovery keep valid state: a revision-bound dispatcher must still apply grants that are current or were re-authorized at the live revision. Section~\ref{sec:kappa-binding} measures this, and we do not map it onto $O_{\mathrm{rt}}$ or $O_{\mathrm{cp}}$.

Reference monitors, lease generations, and optimistic validation are the established principles we build on. The contract adds which state objects and consumers of a stateful AI runtime they must bind, which identity each consumer verifies, and the requirement that exclusion and retention hold together.

\section{Admissible Witnesses: Evidence That a Contract Holds}
\label{sec:oracles}

A contract says what must hold at a consumer. This section asks how we can know that it does. Serving tests normally compare the generated tokens of a request with a clean reference run~\cite{batchinvariant,batchinvvllm,numdet}. That signal reads one request's text, whereas the contracts of Section~\ref{sec:contracts} concern which blocks a later request attends and which action it dispatches. The requirements below limit which evidence may decide a predicate. They do not define correctness, which the contract already does. We state the conditions a witness must meet, show that each common signal certifies one predicate and misjudges the others, and validate the verifier built on the evidence that remains.

\subsection{What makes a witness admissible}
\label{sec:oracle-admissibility}

An observation procedure $O$ that evaluates predicate $P_j$ at consumer $c$ is admissible for contract $\mathcal C$ when
\begin{equation}
\begin{split}
\op{Adm}(O,\mathcal C,c)\ \Leftrightarrow\ &\op{Identity}\wedge\op{SemanticSufficiency}\\
&\wedge\op{Independence}.
\end{split}
\label{eq:admissibility}
\end{equation}
$\op{Identity}$ binds the witness to the episode, the allocation generation, and the consuming request, so the evidence describes the request that read the state. $\op{SemanticSufficiency}$ requires that agreement of the observation entail $P_j$, not merely that a correlated signal match. $\op{Independence}$ requires that the witness not come from harness labels or from the index counters under test~\cite{testoracle,testoracle2,metamorphic,difftest,difftest2}; it rules out circular evidence that defines its own verdict.

Independence here means non-circularity, which is weaker than independent ground truth. Four notions stay apart. \emph{Non-circular evidence} means the witness does not take its verdict from the signal under test, and it is all that admissibility requires. \emph{Independent parsing} means a second reader derives verdicts from the same recorded events, as the two blind language-model parsers of Section~\ref{sec:oracle-validation} do. It tests whether the specification can be read unambiguously. \emph{Independent instrumentation} would mean a separately built recorder. We do not have one: the reference verdicts and the offline verifier read the same recorded event stream, so they share the failure modes of the instrumentation that produced it. \emph{Independent physical ground truth} would require observing device memory itself. Only the Tier-2b digests do, and only where they were collected. We call a witness that meets the three conditions \emph{admissible}.

\begin{proposition}[Evidence-bound verification]
\label{prop:evidence}
Let $O$ be an observation procedure for predicate $P_j$ at consumer $c$. Dropping any one condition of Equation~\ref{eq:admissibility} yields a procedure whose verdict can be disconnected from $P_j$. Without Identity, it may describe a different request. Without Semantic Sufficiency, its agreement need not entail $P_j$. Without Independence, understood as non-circularity, its verdict is fixed by the signal it is scored against. The statement concerns the predicates tested in this paper. It does not characterize all possible observers.
\end{proposition}

\begin{figure*}[!htbp]
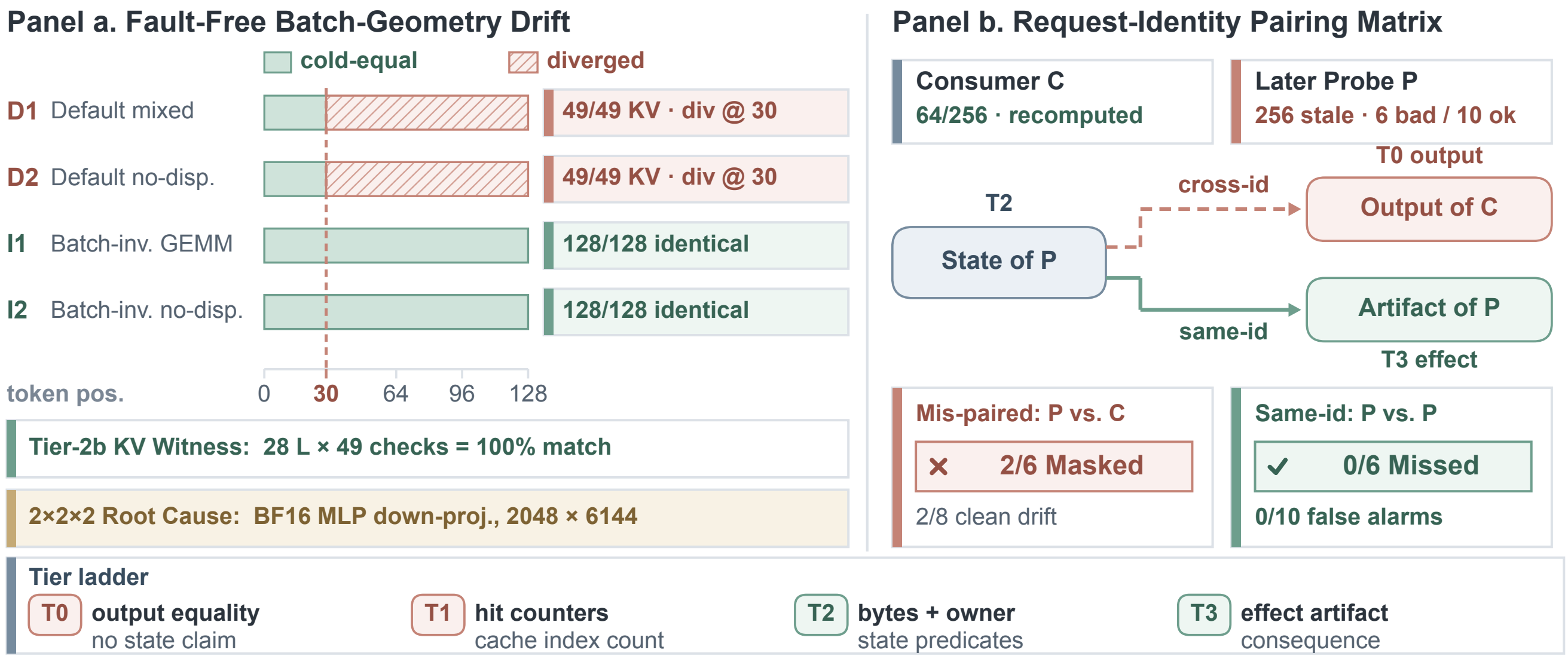

\centering
\PaperFigureTwo{\textwidth}
\caption{\textbf{Output equality fails as a recovery check in two ways.} Panel a: in four fault-free 128-token runs with no transport abort, the two default-arithmetic mixed-batch runs keep identical attended KV digests over 49 forward checks and 28 layers, yet diverge from the cold reference at token 30. Batch-invariant GEMM controls match at all 128 positions. Panel b: pairing consumer C's output with probe P's corrupted state masks $2/6$ corruptions. Pairing P's state with P's own persisted artifact detects $6/6$ with $0/10$ false alarms.}
\label{fig:oracle-counterexample}
\end{figure*}

\noindent\emph{Justification.} The evidence is uneven across the three conditions. Identity and Semantic Sufficiency each have a measured counterexample on the fault corpus of Section~\ref{sec:tiering}. Independence is justified by excluding circular evidence, and no comparable experiment backs it. Output equality is bound to the request and independent of harness labels, yet it is not sufficient: it rejects a correct run under BF16 batching and accepts an exclusion failure (Figure~\ref{fig:oracle-counterexample}a and Table~\ref{tab:tiering}). A hit counter fails Semantic Sufficiency for a similar reason, since it certifies an index entry and not the bytes behind it. Judging the probe's state by the original consumer's output fails Identity, because that consumer never attended the stale suffix, and it accepts $2/6$ corruptions (Figure~\ref{fig:oracle-counterexample}b). For Independence, a counter scored against an exclusion truth that the same counter defines agrees with itself whatever the state is, which is why Table~\ref{tab:tiering} leaves that pairing blank. This is an argument from construction, not a measurement. Nothing here shows that the three conditions are jointly sufficient, and the proposition says nothing about predicates we did not evaluate.

Admissible witnesses differ in strength by what they depend on (Table~\ref{tab:tiering}), and they support different claims. Tier~2a reads integer scheduler state (completion bounds, allocation generations, host and device step counters) and establishes trace conformance, holding only if that instrumentation reports the scheduler's state accurately. Tier~2b hashes the attended KV pages and establishes byte-level agreement, but only for the requests, layers, and runs where it was collected. Tier~3 re-reads the persisted tool artifact and establishes the generated call that was observed and not a backend effect, and effect predicates need it because the conversation identifier carries no principal isolation~\cite{revoked}.

\begin{lemma}[Observation criterion for paired Tier-2a checks]
\label{prop:tier2a}
Assume a scheduler that increments $g$ on every reallocation, accurate completion tracking $V(x)$, and readable host and device step counters. Then a paired Tier-2a check rejects if and only if the recorded metadata show that an admitted prefix position lies outside $V(x)$, that a report's generation differs from the live $g$, or that the host and device step counters differ. Under the same assumptions, these three conditions are violations of production completion, of the generation match, and of $\op{SeamOK}$ in its recorded-counter form. The lemma does not cover token identity, other causes of a corrupted continuation, the contents of device memory, or consumer-time validity as a whole.
\end{lemma}

\noindent\emph{Proof sketch.} The first condition negates $[0,r)\subseteq V(x)$ of Equation~\ref{eq:prefix}, the second negates the generation match of Section~\ref{sec:state-validity}, and the third negates $\op{SeamOK}(e,r)$, which Equation~\ref{eq:prefix} defines as agreement of the recorded host and device positions. Under the stated assumptions the recorded integers are the quantities these predicates mention, so the check rejects exactly when one of the three fails in the record. The check reads integers, so its verdict does not depend on BF16 batching order. It tests the consistency of reported metadata. It does not show that the bookkeeping matches device memory, which the Tier~2b digests check separately.

\subsection{Why output equality and hit counters fall short}
\label{sec:tiering}

\subhead{Numerical drift alone changes outputs}
A fault-free control varies batch schedule and kernel arithmetic with no transport fault and no recovery transition. Isolated transfers match the cold reference. Concurrent mixed-batch runs diverge at token 30 although the SHA-256 digests of the attended cache are identical in all 28 layers (Figure~\ref{fig:oracle-counterexample}a). Enabling \code{VLLM\_BATCH\_INVARIANT=1} restores agreement at all 128 positions. A per-projection replay and an 18-run factorial localize the drift to batch-dependent BF16 accumulation in the $2048\times6144$ MLP down-projections, within $3.095\times10^{-15}$ of exact arithmetic (Appendix~\ref{app:numeric}). Output equality can therefore reject a run whose recovery was correct. The result does not say that BF16 is inaccurate: a fault-free run can fail the usual acceptance test, so that test cannot serve as the oracle for recovery security.

\subhead{Under real faults each signal answers a different question}
A 100-run benchmark with injected faults, ten runs per cell, crosses three source trees with \code{fail} and \code{recompute} and adds the four-cell eviction-by-rewind matrix of Section~\ref{sec:conjunctive}. We score every signal against reference values of three predicates (Table~\ref{tab:tiering}). \emph{Exclusion} asks whether the probe was admitted onto positions that were never written. \emph{Coherence} asks whether the continuation carries the consumer's own value. The position condition $\op{SeamOK}$ of Equation~\ref{eq:prefix} is what the paired step counters read. \emph{Effect} asks whether the re-read tool artifact carries a schema-valid call that the consumer was not authorized to make. None of the 100 runs carries such a call. Its disagreements differ in kind from the fault-free drift above, because the position split damages the continuation without any numerical cause.

Output equality tracks coherence in all 100 runs but is wrong about exclusion in 30. In the rewind-only cell it matches the cold output while 128 unwritten positions are admitted, and in the two eviction-only cells it reports a difference although exclusion holds. Hit counters make the converse error on coherence, also in 30 runs: 128 admitted tokens in the eviction-only cells pass a corrupted continuation, and 256 in the rewind-only cell flag a coherent one. The 50 and 40 in the Effect column count runs in which output equality and hit counters flag a difference although no unauthorized call was generated. The state predicates do fail in these runs, so the alarms are correct about state and uninformative about effect, and they are not BF16 false alarms. No cheap signal decides all three predicates, so each needs its own admissible witness.

\begin{table*}[!htbp]
\centering
\caption{\textbf{Each signal agrees with the predicate it certifies and misjudges others, whereas Tier-2 metadata and Tier-3 artifacts decide the predicates they target.} Hundred fault-injected runs. Each entry counts runs where the signal contradicts the reference value of the column's predicate, taken from the harness's scheduler events and re-read artifacts; a bold zero marks the predicate the signal is meant to certify. A dash marks a pairing that is circular or not meaningful: the hit count is itself the admission measure that defines the exclusion truth, Tier~2a does not observe effects, and Tier~3 defines the effect truth. The device counter is absent from this corpus, so the paired host and device result comes from a separate 14-run cohort.}
\label{tab:tiering}
\scriptsize
\setlength{\tabcolsep}{4.0pt}
\renewcommand{\arraystretch}{0.96}
\begin{tabularx}{\textwidth}{@{}>{\bfseries}c >{\raggedright\arraybackslash}p{0.245\textwidth} >{\raggedright\arraybackslash}X c c c r@{}}
\toprule
Tier & Signal & What it certifies and its limit & Excl. & Coh. & Effect$^{\dagger}$ & Cost\\
\midrule
0 & Output equality with the cold reference & Whether the text matches. Cannot separate a position split from an exclusion failure & 30 & \textbf{0} & 50 & No GPU\\
1 & Cache-hit counter & An index entry, never the written bytes & --- & 30 & 40 & No GPU\\
2a & Allocation identity, generation, completion & Exclusion, read from block metadata at report consumption. Silent on position drift & \textbf{0} & 30 & --- & $25.7\,\mu$s/step\\
2a & Host and device step counters, paired & Coherence. On a separate 14-run cohort it flags all 8 split runs (host 128, device 288) and none of 6 coherent runs & --- & \textbf{0/14} & --- & $25.7\,\mu$s/step\\
2b & SHA-256 of attended KV pages & The bytes attention read. Not collected on this corpus & n/a & n/a & n/a & 0.42 ms/layer\\
3 & Parser output and re-read tool artifact & A generated unauthorized call, which no state signal establishes. Not a backend effect & --- & --- & ref. & One episode\\
\bottomrule
\end{tabularx}
\par\smallskip
{\scriptsize $^{\dagger}$Incorrect if used as an effect verdict. All 100 runs carry no unauthorized call.}
\end{table*}

\subhead{The witness must belong to the consuming request}
Deterministic arithmetic still fails if the test inspects the wrong request. After a failed load the original consumer $C$ receives 64 valid prefix tokens and recomputes its suffix, while a later probe $P$ that matches the same prefix hits all 256 indexed tokens with 192 unwritten. Scheduler logs split the 16 packaged runs of Appendix~\ref{app:pairing} into 6 corrupted runs and 10 clean runs. Pairing $P$'s state predicate with $P$'s own persisted artifact misses none of the 6 corruptions and raises no alarm on the 10 clean runs. Pairing the same state with $C$'s output accepts 2 of the 6, because $C$ never attended the stale suffix (Figure~\ref{fig:oracle-counterexample}b). This is the Identity condition of Equation~\ref{eq:admissibility}.

\subsection{Validating the verifier}
\label{sec:oracle-validation}
\label{sec:rejudge}
\label{sec:false-negative}

We implement the Tier-2a and Tier-3 witnesses as a standalone, GPU-free offline trace verifier. It reads frozen scheduler traces and persisted artifacts with harness labels stripped, and it fails closed when an admissible witness is missing. Attended-KV digests are not its inputs. A separate online Tier-2b audit collects them (Section~\ref{sec:witness-cost}). We test the verifier on the 42-cell frozen GPU matrix of Appendix~\ref{app:audit}. Of the 42 cells, 39 run to completion and 3 block-reuse cells end early in a cuBLAS execution failure after failing their allocation-identity obligation. The verifier agrees with the reference verdicts on $42/42$ scenario verdicts, and a second H20 host with disjoint GPU UUIDs reproduces the result.

At the level of individual obligations, the 210 obligation slots yield 196 applicable and 162 comparable obligations (Appendix~\ref{app:audit}). The verifier agrees with the reference on $159/162$ of the comparable obligations, with $3/149$ conservative false alarms and $0/13$ false witnesses, and the second host reproduces the $159/162$ agreement. All three disagreements are conservative. The reference withholds the timestamp-ordering check after the early abort, while the raw log shows load $98<$ abort $106<$ reallocation $124<$ report consumption $127$. Two blind cross-vendor language-model parsers agree with the reference on $158/158$ obligations. That result tests whether the specification can be read unambiguously, not whether it is true of the hardware.

The reference and the verifier read the same recorded event stream, so the $42/42$ agreement is scenario-level conformance to the specified trace predicates. It is not evidence that the instrumentation is correct or that device state matches the metadata. Byte-level evidence comes only from the Tier-2b audit where it was collected, and effect evidence only from persisted Tier-3 artifacts. Unlike output comparison, the verdicts survive a change of arithmetic. Re-running every output-flagged GPU cell under \code{VLLM\_BATCH\_INVARIANT=1} with fresh within-mode references and fixed schedule, connector, and fault, the Tier-2 verdicts for prefix exclusion, owner disposition, and suffix eviction do not change between modes while the generated tokens do (Appendix~\ref{app:rejudge}).

\section{Boundary Repairs and Evaluation Method}
\label{sec:method}

\begin{figure*}[!htbp]
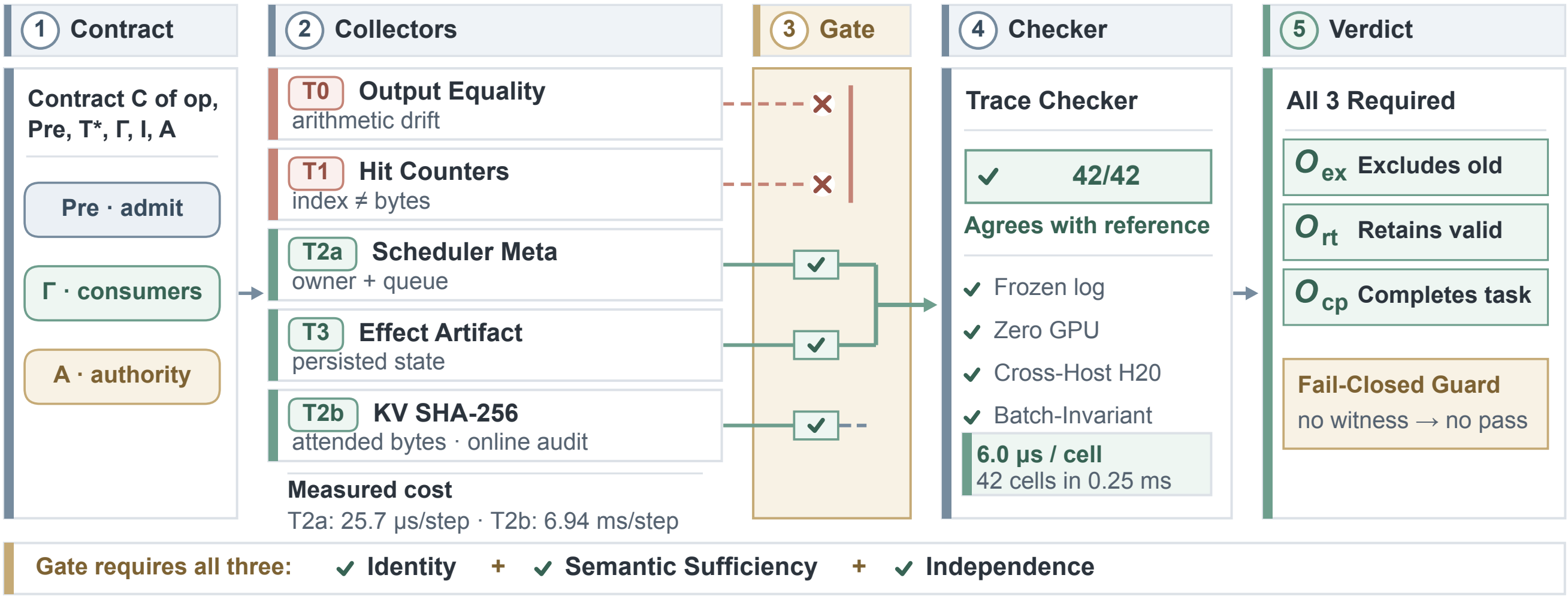

\centering
\PaperFigureThree{\textwidth}
\caption{\textbf{The verifier evaluates consumer-bound recovery predicates using admissible recorded evidence; enforcement remains in the target runtimes.} Contracts determine which predicates must be decided. An admissibility gate rejects Tier~0 (output equality) and Tier~1 (hit counters) for state predicates and admits Tier-2a scheduler metadata and persisted Tier-3 artifacts. These two feed the offline trace verifier, which reads frozen traces, uses no GPU tensors, and takes $6.0\,\mu$s per cell. Tier-2b attended-KV digests are admissible evidence but are collected by a separate online audit (Section~\ref{sec:witness-cost}), not by the offline verifier. Repairs are separate code in each target system. The three obligations in the verdict panel, $O_{\mathrm{ex}}$, $O_{\mathrm{rt}}$, and $O_{\mathrm{cp}}$, apply to state-recovery contracts. Authority-freshness predicates are Equations~\ref{eq:revocation} and~\ref{eq:staleness}.}
\label{fig:witness-architecture}
\end{figure*}

\subsection{Repairs enforce the contract at the consumer}
\label{sec:exposure}

Repairs are the only part of the method that changes run-time behavior. A repair is adequate only when it restores the consumer-side conditions the contract requires without giving up what remains valid. For a state-recovery contract, this means recovery adequacy: exclusion together with retention and completion. For the dispatcher, it means refusing stale grants while still applying current and re-authorized ones. Clearing every cache or ending every session meets the first requirement and fails the second. Repairs therefore belong at the boundary that can enforce the condition, and their mechanisms differ because the consumers verify different identities (Table~\ref{tab:repairs}). In each case the consumer stops relying on an upstream record alone and validates an identity of its own. The Tier-2a hooks and the offline verifier of Section~\ref{sec:oracles} observe whether the postcondition holds and block nothing (Figure~\ref{fig:witness-architecture}), and a repair that passes them on the tested paths is not thereby shown correct on others.

\begin{table}[!htbp]
\centering
\caption{\textbf{Boundary-specific repairs enforce the consumer's postcondition.} The identity column gives the value the consumer verifies, from Section~\ref{sec:identity-spaces}.}
\label{tab:repairs}
\footnotesize
\setlength{\tabcolsep}{3.0pt}
\renewcommand{\arraystretch}{0.95}
\begin{tabularx}{\columnwidth}{@{}>{\raggedright\arraybackslash}p{0.17\columnwidth} >{\raggedright\arraybackslash}X >{\raggedright\arraybackslash}p{0.19\columnwidth}@{}}
\toprule
System & Repair at the consumer & Identity\\
\midrule
ChatGLM2 & Rebuild the prompt and cache from the retained prefix after a deletion & $\mathcal{T}$\\
Google ADK & Export only events from the current rewind epoch & $\mathcal{T}$\\
vLLM & Evict unwritten hashes by allocation, settle \code{PREEMPTED} owners, rewind the device position & $\mathcal{T}$, $g$\\
SGLang & Send KV state to decode only after intake accepts the request & $\mathcal{T}$\\
DSH & Dispatch a grant only if its revision equals the live revision & $\kappa$\\
\bottomrule
\end{tabularx}
\end{table}

\subsection{Evidence modes and cost}
\label{sec:witness-cost}

Three kinds of evidence appear in the evaluation, and each answers a different question. \emph{Live exploitation} runs native lifecycle transitions end to end and asks whether a failure is reachable. \emph{Deterministic conformance} covers the 42-cell serving matrix, 300 dispatcher boundary instances, and 7 archived provider replays, and asks whether a tree or dispatcher satisfies the contract on specified cases. \emph{Schedule sensitivity} sweeps 4,000 seeded interleavings in a controlled dispatcher model and asks how sensitive the dispatcher is to timing. Conformance and sweep counts are not field prevalence. Every cohort pins a commit, checkpoint, and hardware stack (Appendix~\ref{app:gpu}) and holds token composition, positional layout, decoding policy, and seeds fixed across matched arms. Appendix~\ref{app:coverage} registers all 29 coverage units, each pairing a contract, consumer, transition, and configuration, with unpooled denominators, and the DSH episodes keep a separate denominator.

Online observation must not disturb the scheduler. On the pinned H20 stack, with eight concurrent requests and 152 live GPU blocks over $9.6\times10^{6}$ timed steps in nine runs, the Tier-2a hooks cost $25.7\,\mu$s of host CPU per step (p95 $29.7\,\mu$s) with no GPU kernel and no tensor read. The optional Tier-2b byte audit costs $0.42$\,ms per step for one layer and $6.94$\,ms for all 28. Auditing one canary layer adds $4.3\%$ latency and retains $96\%$ of throughput, and hashing all 28 layers raises mean latency by $26.3\%$.

\begin{table}[!htbp]
\centering
\caption{\textbf{Operator workarounds close fewer paths than boundary repairs.} Closed means every path exercised in this paper on that boundary is removed, and open means one survives. The repair row is measured, and the other rows follow from the analysis of Sections~\ref{sec:consumers} and~\ref{sec:producers}.}
\label{tab:mitigations}
\footnotesize
\setlength{\tabcolsep}{3.0pt}
\renewcommand{\arraystretch}{0.95}
\begin{tabularx}{\textwidth}{@{}>{\raggedright\arraybackslash}p{0.30\textwidth}
                              >{\raggedright\arraybackslash}p{0.11\textwidth}
                              >{\raggedright\arraybackslash}p{0.09\textwidth}
                              >{\raggedright\arraybackslash}p{0.12\textwidth}
                              >{\raggedright\arraybackslash}X@{}}
\toprule
Mitigation & Prefill & Pool & Cross & Intra\\
\midrule
Disable prefix caching & lost & kept & closed & open\\
Per-tenant \code{cache\_salt} & $K\times$ HBM & kept & closed & open, same tenant\\
Disable external KV tier & local & lost & closed & open\\
Conjunctive repair & kept & kept & closed & closed\\
\bottomrule
\end{tabularx}
\end{table}

Table~\ref{tab:mitigations} compares three coarse workarounds. Disabling prefix caching forfeits shared-prefix reuse; an archived microbenchmark measures $293.62$\,ms for a cold $8192$-token prompt against $30.58$\,ms for an uncached suffix rebuild, a $9.6\times$ penalty. A per-tenant \code{cache\_salt} separates tenants but stores the prefix once per tenant and leaves intra-tenant paths open. Disabling the external tier gives up cross-node pooling and does not stop corrupted local continuation. Repair at the consuming boundary keeps both reuse and pooling.

\section{Single-Session Consumers: Why Clearing State Is Not Recovery}
\label{sec:consumers}

Two single-session systems show why recovery adequacy requires exclusion, retention, and completion together: in each, a user-visible record is updated correctly while a derived representation that another consumer reads is not.

\subsection{ChatGLM2: deletion prunes the transcript but not the KV cache}

We replay the branch-deletion and generation callbacks of an archived ChatGLM2 deployment~\cite{space} on a pinned FP16 snapshot (Transformers 4.30.2). Deleting a turn prunes the visible transcript and the history list but leaves \code{past\_key\_values} intact, and across nine low-temperature conditions the forward-pass tensor digests match the deleted branch in all $9/9$ runs. Clearing the cache inside the callback removes the stale tensors and also strips every retained turn in $9/9$ conditions, because the wrapper keeps \code{return\_past\_key\_values} enabled and its prompt constructor then encodes only the current round.

\begin{figure}[!htbp]
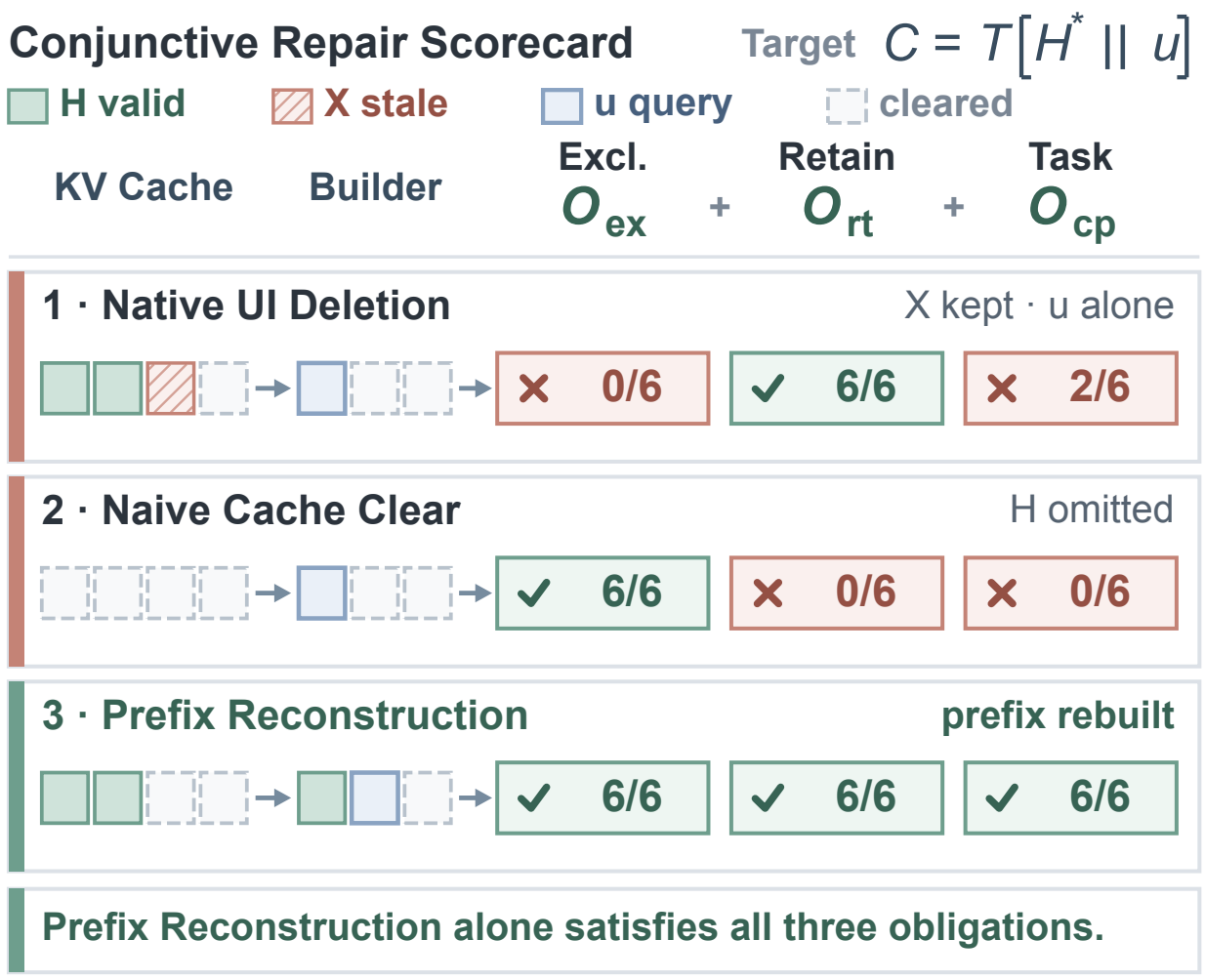

\centering
\PaperFigureFour{3.39in}
\caption{\textbf{A repair that satisfies exclusion alone breaks retention.} Six matched conditions, three history-dependent tasks, two deletion depths. Native deletion keeps the revoked branch $X$ in the cache. Naive clearing removes $X$ together with retained history $H$. Prefix reconstruction keeps $C_{:r}$ and rebuilds $C_{r:}$, and satisfies exclusion $O_{\mathrm{ex}}$, retention $O_{\mathrm{rt}}$, and completion $O_{\mathrm{cp}}$ in $6/6$ runs.}
\label{fig:conjunctive-repair}
\end{figure}

Figure~\ref{fig:conjunctive-repair} compares five recovery regimes over six matched conditions, with three history-dependent tasks and two deletion depths. Native deletion violates exclusion in $6/6$ conditions yet still answers $2/6$ tasks from surviving context, so a task-accuracy metric hides the leak. Naive clearing satisfies exclusion in $6/6$ and drops retention and completion to $0/6$. Reconstructing the prompt and cache from the retained prefix satisfies all three in $6/6$ runs and matches full recomputation: a correct repair rebuilds the state that the canonical retained history implies (Section~\ref{sec:contract-def}).

\subsection{Google ADK: a filtered view does not bind the exporter}
\label{sec:pending-tail}

Google ADK 2.13.0 keeps a filtered active-event view after a session rewind, but its in-memory long-term exporter scans the unfiltered raw event log~\cite{adkissue}. The filtered view is correct for its own consumer, but the exporter does not inherit its recovery semantics. We replay native rewind, memory ingestion, and retrieval over nine branched histories. The unmodified exporter satisfies the post-rewind contract in $3/9$ histories and exports rewound turns in the other six. Filtering exports by rewind epoch, applying upstream commit \code{3640e7d0c74b}, or exporting from reconstructed active events gives $9/9$ compliance across 36 conditions. Three pre-rewind controls mark its limit: turns an external store ingested before the rewind persist until an explicit cross-store revocation runs.

\subsection{Cline: a negative control}

Cline's checkpoint restore serves as a negative control and not as a further vulnerability. Its five controls (restore, repeat restore, late append, compaction, and a native effect) all behaved correctly under the tested configuration (Appendix~\ref{app:coverage}). It is therefore counted among the 21 boundary controls of the coverage register and not among the seven mediation failures. The control shows only that the evaluation does not flag every rewind interface it touches.

\section{Disaggregated KV Serving: From an Interrupted Load to a Wrong-Target Call}
\label{sec:producers}

An incomplete KV load becomes a security problem once the pool is shared. This section follows one fault from the unwritten blocks it leaves indexed to a second tenant's tool call. The consumer is the scheduler's prefix-admission path and the runner that continues from the admitted position, and both rely on the prefix index and on the scheduler's host-side counters. The upstream event is a failed or canceled load. Sections~\ref{sec:kv-effect} and~\ref{sec:kv-policy} show that the consumer still admits the unwritten state, and Section~\ref{sec:state-obligations} shows why a published patch closes only part of the gap. Section~\ref{sec:conjunctive} separates the two defects that a repair must fix, Section~\ref{sec:pav} follows the corrupted state into a generated tool call, and Section~\ref{sec:state-sufficiency} states what the chain establishes. All serving experiments use vLLM 0.33.0rc1 with LMCache V3 and the \code{LocalCPU} connector.

\subsection{An aborted load leaves unwritten blocks indexed}
\label{sec:kv-effect}

An interrupted load can leave unwritten blocks indexed because the index records that a block was registered and not that it was written. A priming request populates a 256-token external prefix, and a consumer load then aborts at token 64. Connector backpressure, remote disconnect, pool eviction, or client cancellation can cause such an abort. The prefix index still advertises all 256 tokens and the connector has recorded the failure, but the producer completed only 64. An index entry says that a block with this hash was registered, not that its bytes can be read, so the 192 suffix tokens over 12 GPU blocks stay matchable. We verify the outcome from persisted Tier-3 artifacts. In all six stale-KV runs on the unpatched tree and on upstream PR 61769, the later probe emits the degenerate repetition \code{[525]}$\times 8$ instead of \code{" 1, 2, 3"}. The three columns of Panel~A of Table~\ref{tab:kv-policy} score different outcomes. Our allocation-aware repair, which admits 64 valid tokens, produces the expected probe artifact in all three runs and eliminates the recorded stale-state effect in all three runs, but only one of the three consumer outputs matches the cold reference token for token. The cache-only control, which admits the full 256-token prefix, matches the cold reference in all three runs. Artifact correctness, cold-reference output equality, and the stale-state effect are therefore distinct measurements, and a repair that restores the artifact does not by itself restore every consumer output; the archive does not establish the cause of the two consumer-output mismatches.

\begin{table*}[!htbp]
\centering
\caption{\textbf{The failure policy determines cross-tenant reachability, and a complete repair needs allocation-aware eviction $E$ with device position rewind $D$.} One pinned vLLM and LMCache stack, tenants sharing a 256-token prefix; Adm is prefix tokens admitted without computation. Panel A end-to-end cohort (Art: probe artifact matches the cold reference; Cons: consumer output matches the cold reference token for token; Eff: recorded stale-state effect; the three are distinct measurements and are not pooled). Panel B: trees crossed with \code{fail} and \code{recompute}. Panel C: the $2\times2$ factorial of $E$ and $D$.}
\label{tab:kv-policy}
\scriptsize
\renewcommand{\arraystretch}{0.95}
\begin{minipage}[t]{0.305\textwidth}
\textbf{Panel A. End-to-end contamination}\par\vspace{2pt}
\setlength{\tabcolsep}{2.4pt}
\begin{tabular*}{\linewidth}{@{\extracolsep{\fill}}l l r r c c c@{}}
\toprule
Arm & Tree & \textit{n} & Adm & Art & Cons & Eff\\
\midrule
Stale KV & Upstr. & 4 & 256 & Degr & 1/4 & 4/4\\
Stale KV & PR & 2 & 256 & Degr & 1/2 & 2/2\\
Repaired & Cand. & 3 & 64 & OK & 1/3 & 0/3\\
Cache-only & Upstr. & 3 & 256 & OK & 3/3 & 0/3\\
No-fault & Upstr. & 2 & 256 & OK & 2/2 & 0/2\\
Fail-closed & Upstr. & 2 & 64 & OK & 0/2 & 0/2\\
\bottomrule
\end{tabular*}
\end{minipage}\hspace{7pt}%
\begin{minipage}[t]{0.425\textwidth}
\textbf{Panel B. Policy reachability across trees}\par\vspace{2pt}
\setlength{\tabcolsep}{2.2pt}
\begin{tabular*}{\linewidth}{@{\extracolsep{\fill}}l l r r l c c@{}}
\toprule
Tree & Policy & \textit{n} & Adm & Owner status & Tool & Cont.\\
\midrule
Upstream & \code{fail} & 10 & 128 & \code{FINISHED\_ERROR} & OK & clean\\
Upstream & \code{recomp.} & 10 & 256 & \code{LENGTH\_CAPPED} & Degr & dirty\\
PR 61769 & \code{fail} & 10 & 128 & \code{FINISHED\_ERROR} & OK & clean\\
PR 61769 & \code{recomp.} & 10 & 256 & \code{LENGTH\_CAPPED} & Degr & dirty\\
Alloc.-aware & \code{fail} & 10 & 128 & \code{FINISHED\_ERROR} & OK & clean\\
Alloc.-aware & \code{recomp.} & 10 & 128 & \code{LENGTH\_CAPPED} & Degr & split\\
\midrule
Order ctrl. & \code{recomp.} & 1 & 256 & \code{STOPPED} & OK & clean\\
Cache ctrl. & \code{recomp.} & 1 & 272 & local hit & OK & clean\\
\bottomrule
\end{tabular*}
\end{minipage}\hspace{7pt}%
\begin{minipage}[t]{0.215\textwidth}
\textbf{Panel C. Factorial repair}\par\vspace{2pt}
\setlength{\tabcolsep}{2.4pt}
\begin{tabular*}{\linewidth}{@{\extracolsep{\fill}}c c r r c c@{}}
\toprule
$E$ & $D$ & \textit{n} & Adm & Sound & Corr.\\
\midrule
$\times$ & $\times$ & 10 & 256 & none & none\\
$\pmb{\checkmark}$ & $\times$ & 10 & 128 & $\pmb{\checkmark}$ & none\\
$\times$ & $\pmb{\checkmark}$ & 10 & 256 & none & $\pmb{\checkmark}$\\
$\pmb{\checkmark}$ & $\pmb{\checkmark}$ & 10 & 128 & $\pmb{\checkmark}$ & $\pmb{\checkmark}$\\
\bottomrule
\end{tabular*}
\par\vspace{2pt}
{\scriptsize $E$ bounds admission, $D$ rewinds the GPU position. Sound: admission is bounded. Corr.: continuation is correct.}
\end{minipage}
\end{table*}

\subsection{Policy and lifecycle decide who can reach the blocks}
\label{sec:kv-policy}

Whether a later request reaches the unwritten blocks depends on how the scheduler settles the failure report: which owner it maps the report to, what it evicts, and which operator policy it applies. vLLM handles an external load failure under \code{fail} or \code{recompute}. ALPHA and BETA share a 256-token system and tool-schema prefix and hold private payloads, and ALPHA's load aborts at token 128, leaving the upper 128 tokens unwritten before BETA arrives. Under \code{fail}, ALPHA ends in \code{FINISHED\_ERROR}, the reported blocks are evicted, and BETA is admitted onto the 128 valid tokens with a clean tool call in $30/30$ runs. Under \code{recompute}, the scheduler rewinds ALPHA's host position counter from 288 to 128 and leaves the prefix hashed. BETA then receives all 256 shared-prefix tokens with 128 unwritten and produces corrupted tool output in $20/20$ runs across upstream and PR 61769 (Panel B). Settlement presumes that the owner is still running or waiting on the remote load, but a scheduler under memory pressure can move it to \code{PREEMPTED} before the report arrives.

Each cell of Panels B and C counts ten runs that produced an observation, and the 100 runs in those cells come from one matrix of 126 launches. In 25 launches (20 under \code{recompute}, 5 under \code{fail}) request ALPHA did not finish within the harness's 80-step liveness budget and the process exited with code 137 (21) or 134 (4), and one observed launch carrying no cell label is excluded. The ledger does not establish why ALPHA failed to terminate, so the results rest on the 100 observed runs.

An unprivileged client controls when its own request is canceled, and connector faults abort loads without any client action. On the unpatched and backported trees, the stale hashes stay indexed from the moment the scheduler consumes the failure report until the later probe returns. The median is $204.15$\,ms in the seven lifecycle runs of Figure~\ref{fig:block-lifecycle}, against $11.61$\,ms for the connector report and $0.38$\,ms for settlement. An attacker therefore needs no cycle-accurate race.

\subsection{Why the upstream patch is incomplete}
\label{sec:state-obligations}
\label{sec:current-coordination}

Public trackers already describe parts of this failure and one proposed fix: vLLM issue 61768 and PR 61769 cover an abort that precedes failure processing, issue 61905 covers late reports after block reuse, and issue 61912 describes the preempted-owner gap~\cite{vllmrace,vllmfix,vllmreuse,vllmpreempt}. Those reports establish the failure facts. They do not say whether a patch establishes the condition a later consumer depends on. That is the question the contract answers, so we treat the patch comparison as a test of the method and not a contest between patches. In short, PR 61769 covers the direct-abort path and not every preemption and recycled-block path we tested. We pin vLLM main (\code{b8c04dd3}) and the head of PR 61769 (\code{8ee68eb1}) and run the two trees and our candidate through 14 scheduler cases each, 42 in all. Seven cases are triggered by a fault, and seven are live-owner, successful-load, and consumer-role controls that every tree passes. The scheduler and cache manager are the unmodified pinned source and the connector and runner outputs are deterministic mocks, so the matrix tests state transitions and not GPU bytes. Main satisfies $0/7$ of the triggered predicates and PR 61769 satisfies $2/7$. Our allocation-generation and preempted-settlement repair satisfies $7/7$ with all seven controls intact, so a valid prefix stays reusable (Table~\ref{tab:repair-tree}).

The two predicates that PR 61769 satisfies are the direct-abort cases, one per policy. The handler finds the running owner, evicts the failed suffix, and bounds admission to the 48 valid tokens, where main readmits 112. The five it fails form two groups, each an unmet condition. Three involve a \code{PREEMPTED} owner, through a reset under each policy and through memory-pressure preemption. The handler scans only the \code{running} and \code{remote-waiting} owners, finds no record of the \code{PREEMPTED} one, evicts nothing, and never applies the failure policy, so a later request is again admitted onto 112 tokens. This is the path the live runs take: the owner is \code{PREEMPTED} when the report is consumed (Figure~\ref{fig:block-lifecycle}), and PR 61769 admits all 256 tokens in both of its live stale-KV runs (Panel A). The other two involve a delayed report that arrives after the scheduler has freed and reallocated the block, and a report that names only a bare block identifier fails or rewinds an unrelated new owner. 

Eviction must therefore follow the allocation that produced the report and not queue membership, settlement must cover owners outside the running queues, and a report may act only on the block instance it was issued for. Our repair keeps each scheduled output's allocation records across owner lifecycle changes, stamps physical blocks with allocation epochs, and settles the owner under the operator's policy. A paired \code{continue}$\rightarrow$\code{pass} ablation shows that the stale-output guard is armed in $6/6$ cases, so the gap is not a missing guard. It discards 16 stale tokens on the baseline and commits the stale token under \code{fail} once removed.

Because the PR head and main differ in base commit, the comparison is between trees. The recycled-block and pressure-preemption predicates run in a producer-role configuration with a synchronous-loading mock, and their reachability through a deployed connector is not established. The repair assumes ordered, one-to-one binding of synchronous reports to scheduled outputs, keeps legacy handling for unknown or asynchronous reports, and removes the unsafe readmission after a reset without changing the reset exception. A later community pull request adds preempted-owner settlement to the direct-abort fix and reports the same two open conditions, corroborating the gap the contract names~\cite{vllmpreemptfix}. Both upstream pull requests were closed without merging, PR 61769 on November 12 and the follow-up on November 13, 2026, so neither is a deployed official patch.

\subsection{Why one repair is not enough: eviction and device rewind}
\label{sec:conjunctive}

The recompute path hides two defects that need separate repairs. Upstream vLLM rewinds the host token counter from 288 to 128 but leaves the GPU position buffer at 288, so every later step carries a 160-position offset, and the same missing device rewind is reported upstream as a wrong-output bug in the V2 runner~\cite{vllmrewind}. The factorial of Panel C crosses allocation-aware eviction $E$, which bounds admission to written state, with device position rewind $D$, which restores the continuation. Without either, 256 tokens are admitted and neither condition holds. With $E$ alone, admission is bounded to the 128 valid tokens, so an allocation-level check reports success, yet BETA's continuation stays corrupted in $10/10$ runs. With $D$ alone, the continuation is correct in $10/10$ runs while 128 unwritten positions remain admitted, so an output comparison reports success over a cache that was never written. Only $E\wedge D$ satisfies both conditions in $10/10$ runs, the repaired arm of Table~\ref{tab:pav}. 

In the terms of Equation~\ref{eq:prefix}, eviction restores production completion, $[0,r)\subseteq V(x)$, and device rewind restores execution-position coherence, $\op{SeamOK}(e,r)$. Neither arm implies the other, so each single repair satisfies one condition and passes one natural check while failing the other. On this live path, $E\wedge D$ satisfies both measured conditions and neither repair alone does. We make no claim for other connectors or schedules. Paired Tier-2a counters flag the 160-token split in all $4/4$ eviction-only runs of the 14-run device cohort, the runs in which an allocation check alone reports success.

\subsection{Consequence: a schema-valid call that names the wrong tenant}
\label{sec:pav}

The remaining question is whether the invalid state reaches an action. A corrupted continuation that fails to parse causes only denial of service, whereas the harmful case parses, satisfies the schema, and carries the wrong target. We classify each generated tool-call artifact, written to disk and re-read, into one of three outcomes. A \emph{correct-target call} carries the requesting tenant's own identifier and value. A \emph{wrong-target call} carries another tenant's identifier or an unauthorized payload. \emph{No action} means that parsing or schema validation rejects the artifact. Table~\ref{tab:pav} reports 30 admitted runs, and all 30 artifacts pass schema validation. Without a fault, BETA's session reuses 511 cached tokens that include the shared prefix, and it generates its own correct-target call in $10/10$ runs. When ALPHA's load aborts at token 128 under the native fault, the scheduler admits the stale prefix to BETA's authenticated session, and BETA generates a schema-valid call that names ALPHA in $10/10$ runs, with no correct-target call and no parser rejection. The repair restricts admission to the 128 verified tokens and restores $10/10$ correct-target calls. The stale state thus reaches the target parameter of a generated call and not only its text, although the call is an artifact and no backend executes it.

\begin{table}[!htbp]
\centering
\caption{\textbf{Stale-state admission yields schema-valid tool calls naming the wrong tenant in $10/10$ native-fault runs, and the repair restores $10/10$ correct-target calls.} Ten runs per arm on one pinned harness (identical SHA-256 on both hosts), model, and prompt schedule, \code{recompute} policy, fault at token 128. The native-fault arm ran on one H20 host and the repaired arm on a second; the no-fault baseline gave $10/10$ correct-target calls on each. Adm. counts prompt tokens admitted from cache without computation. Calls are generated and re-read from disk; no backend executes them.}
\label{tab:pav}
\scriptsize
\setlength{\tabcolsep}{5.2pt}
\renewcommand{\arraystretch}{0.98}
\begin{tabularx}{\columnwidth}{@{}X r c c c@{}}
\toprule
Evaluation arm & Adm. & Correct target & Wrong target & No action\\
\midrule
Cold baseline & 511 & 10/10 & 0/10 & 0/10\\
Native fault & 256 & 0/10 & 10/10 & 0/10\\
Repaired & 128 & 10/10 & 0/10 & 0/10\\
\bottomrule
\end{tabularx}
\par\smallskip
{\scriptsize Cold-baseline probes reuse 511 valid cached tokens. Native-fault probes reuse 256 shared-prefix tokens of which 128 were never written. The generated fields are \code{"tenant\_id": "BETA"} with \code{"record\_value": "7, 6, 5"} in all cold and repaired runs, and \code{"tenant\_id": "Alpha"} with a corrupted trailing delimiter in all native-fault runs.}
\end{table}

The 256-token hit has three spans. Positions 0 through 127 hold the prefix that ALPHA's load wrote, including the schema sentence that lists the tenant labels ALPHA, BETA, and GAMMA, with ALPHA first. Positions 128 through 255 are the unwritten rest of the prefix, whose hashes remain indexed, and positions 256 and above are BETA's own body, computed in the current prefill. At the seam the host counts 128 computed tokens while the staged GPU buffer reads 288, so attention applies positional encodings 160 places ahead over a span the load never populated. The emitted label is consistent with the first label of the written list, but we ran no spacer control and do not attribute the label to a span.

The schema types \code{tenant\_id} as an unconstrained string, so a schema-valid call can name any tenant. The emitted label is \code{Alpha}, which differs in case from the prompt's ALPHA, so a case-sensitive backend could treat it differently. The call becomes an unauthorized write only if a backend trusts the model-supplied tenant identifier (Section~\ref{sec:adversary}). ALPHA's private value $1, 2, 3$ appears in none of the 71 re-read BETA artifacts of the harvested cohort, and BETA's own value fills the value field in $10/10$ native-fault runs, so the data show neither attacker-controlled target selection nor a read of ALPHA's private data.

\subsection{What the chain establishes}
\label{sec:state-sufficiency}

SGLang exhibits a related failure at the asynchronous prefill handoff, although its mechanism differs from the vLLM case. Across four intake-rejection runs, two per source tree, the base tree forwards KV state and a substitute token to the decode worker in both runs, and the patched tree withholds the send until intake accepts in both, at the cost of a liveness limit on the rejected request. The claim is bounded to that asynchronous timeout and does not reproduce the vLLM mechanism.

Each body of evidence answers a different question. The scheduler matrix shows which lifecycle conditions a tree covers. The live fault shows that a later consumer reaches the unwritten state. The factorial shows that production completion and execution-position coherence are independent and jointly necessary on the tested path. The tool-call artifacts show that the violation reaches an action parameter.

No single return value, counter, or patch certifies recovery. A recorded connector failure, a \code{FINISHED\_ERROR} status, a bounded hit count, and a matching output can each hold while another condition fails. On the tested path, recovery required five conditions: production over every admitted position, a report resolved to the allocation that produced it, owner settlement outside the running queues, a device position that continues from the host's, and valid state left reusable. Each is checkable on the consuming request with the witnesses of Section~\ref{sec:oracles}.

\section{Authority Freshness: Deferred Grants and Stale Execution}
\label{sec:authority}

The earlier cases ask whether a consumer reads valid state. This one asks whether a dispatcher executes under authority that is still valid, which needs a different identity (Section~\ref{sec:authority-contract}). The consumer is an agent's tool dispatcher and the recorded decision is an approval. When the two are separated in time, a grant that was never revoked passes every static identity check at dispatch and can still overwrite newer workspace state. This is the classical TOCTOU gap, which optimistic concurrency control closes by validating at commit~\cite{occ}. Three conditions separate here. A grant can stay unrevoked without staying valid for the target's current revision. A dispatcher can compare revisions and still lose a commit that lands before its write. Only a comparison inside an atomic check-and-write meets Equation~\ref{eq:staleness}.

\subsection{A live multi-session overwrite}
\label{sec:stale-writer}

We build an end-to-end counterexample on the DSH approval runtime~\cite{dsh,dshruntime} with live model calls and native tools. Session~1 reads a protected file at revision $\kappa$ and queues a write for approval. Session~2 independently obtains approval and commits a newer value at $\kappa+1$, and Session~1's grant is then released while Session~3 reads the file. Four dispositions of the grant (\code{allowed-once}, \code{canceled}, \code{rejected}, \code{serialized}) across two model families give eight episodes, all kept in the denominator. Five reach dispatch, and three stop before $\op{Pre}$ because setup produced no live pending draft (Appendix~\ref{app:agents-sec}). The first three dispositions withhold execution before dispatch and satisfy Equation~\ref{eq:revocation}. Only \code{allowed-once} releases an unrevoked grant whose target has moved, which tests Equation~\ref{eq:staleness} directly. The dispatcher rechecks principal, path, operation type, and permission version, never reads $\kappa$, and overwrites Session~2's newer state in the $1/1$ admitted execution.

\subsection{Revision-bound dispatch}
\label{sec:kappa-binding}
\label{sec:async-authority}

The live execution establishes reachability, and three controlled bodies of evidence with separate denominators test the version check across more interleavings. First, 300 deterministic boundary instances that satisfy $\op{Pre}$ compare the revision-unbound dispatcher with a revision-bound one that enforces Equation~\ref{eq:staleness} (Table~\ref{tab:kappa-binding}): 160 \emph{conflicting} grants overtaken at $\kappa+1$, 40 \emph{current} grants at $\kappa$, and 100 \emph{re-authorized} grants approved again at $\kappa+1$. Second, in seven archived provider-backed replays both dispatchers apply all $7/7$ at their original timestamps, and after a concurrent write the unbound dispatcher reverts newer state in $7/7$ while the bound one blocks all $7/7$.

\begin{table}[!htbp]
\centering
\caption{\textbf{Revision binding blocks $160/160$ conflicting overwrites and applies all $140/140$ valid grants.} Three hundred deterministic boundary instances, with identical grant tuples for both dispatchers. Stale counts silent overwrites of a newer revision $\kappa+1$.}
\label{tab:kappa-binding}
\scriptsize
\setlength{\tabcolsep}{2.6pt}
\renewcommand{\arraystretch}{0.95}
\begin{tabularx}{\columnwidth}{@{}X r r r r r r@{}}
\toprule
& & \multicolumn{3}{c}{\textbf{Unbound}} & \multicolumn{2}{c}{\textbf{Bound}}\\
\cmidrule(lr){3-5}\cmidrule(lr){6-7}
Family & \textit{n} & Stale & Appl. & Ref. & Stale & Appl.\\
\midrule
Conflicting & 160 & 160 & 160 & 0 & \textbf{0} & 0\\
Current & 40 & 0 & 40 & 0 & 0 & \textbf{40}\\
Re-authorized & 100 & 0 & 100 & 0 & 0 & \textbf{100}\\
\bottomrule
\end{tabularx}
\end{table}

Third, a seeded dispatcher model, separate from the live run and the replays, varies approval latency with $1{,}000$ episodes for each of four commit arrivals (\code{immediate}, \code{deferred}, \code{racing}, \code{reordered}). The unbound dispatcher reverts newer writes in all $1{,}000$ \code{deferred} runs and in all $502$ \code{racing} runs where the concurrent commit lands first, $1{,}502$ rollbacks in all, and revision-bound dispatch removes all of them with zero false refusals across the $4{,}000$ episodes. Both dispatchers receive identical grant tuples. These are model episodes, not agent attacks, and they do not establish field prevalence. A fixed timeout cannot substitute for the check, since a short window rejects deliberate review and a long one admits rapid edits. The comparison $\kappa_{\mathrm{grant}}=\kappa_{\mathrm{now}}$ at dispatch resolves causal conflicts whatever the queue delay, and it excludes a concurrent write between check and write only inside the dispatcher's atomic check-and-write section.

\section{Discussion: Recovery as a Consumer-Side Property}
\label{sec:principles}
\label{sec:root-cause-map}

\begin{table}[!htbp]
\centering
\caption{\textbf{Seven interface requirements tie each cross-boundary failure to an observation on a pinned test or to the lifecycle analysis, and five rest on observed failures and repairs.} Observed: a failure and a repair were measured on the named pinned system. Design implication: derived from the lifecycle analysis without a dedicated experiment.}
\label{tab:principles}
\scriptsize
\setlength{\tabcolsep}{2.4pt}
\renewcommand{\arraystretch}{0.92}
\begin{tabularx}{\columnwidth}{@{}>{\raggedright\arraybackslash}p{0.13\columnwidth} X >{\raggedright\arraybackslash}p{0.145\columnwidth}@{}}
\toprule
Interface & Requirement and motivating observation & Basis\\
\midrule
Selection & Pass the post-recovery target $T^*$ to every adapter and verify prefix identity. ChatGLM2 deletion reads removed-branch KV in $9/9$, and a naive clear drops retained history in $6/6$ & Observed, 6 endpoints\\
Publ.\ status & Bind each completion claim to a receipt for the episode. A denied edit is recorded as completed while the receipt logs the denial & Observed, 1 endpoint\\
Ownership & Bind reports to allocation generation $g$ and settle \code{PREEMPTED} owners. Upstream admits 112 prefix tokens where 48 are valid, and PR 61769 leaves \code{PREEMPTED} owners unresolved & Observed, 7 endpoints\\
Retirement & Quarantine blocks until in-flight DMA and kernel readers retire. Attribution alone cannot fence an asynchronous read & Design implication\\
Publ.\ gating & Re-verify assignment and prefill acceptance before transfer and decode handoff. SGLang forwards KV state and a substitute token after intake rejection & Observed, 1 endpoint\\
Action auth. & Revalidate principal, target, operation, permission version, and live $\kappa$. Binding blocks $160/160$ conflicts and $1{,}502/1{,}502$ sweep conflicts & Observed, 1 live ep., 300 inst., 7/7 replays\\
Ack. & Distinguish $t_{\mathrm{cancel}}$, $t_{\mathrm{settle}}$, and $t_{\mathrm{retire}}$ in the API. One boolean flag creates the preemption and reuse windows & Design implication\\
\bottomrule
\end{tabularx}
\end{table}

\subhead{Recovery is decided at the consumer}
Persistent caches, long-term memory, asynchronous tools, checkpoint restoration, and resource reuse each move state beyond the request that created it. A logical cancellation, a deleted history entry, or a completed approval therefore decides less about later execution than the component that recorded it assumes. Proposition~\ref{prop:ack} states the consequence: an acknowledgment describes the moment it was issued, and in the evaluated runtimes it held while a predicate the consumer needed failed.

The failures look different across systems, yet each reproduced failure reduces to one of three missing bindings at a consuming boundary: a representation used after revocation, a report unbound to an allocation generation, and a dispatch blind to the target's revision. Table~\ref{tab:sufficiency-bridge} lists the six failures with, for each, the boundary it crosses, the missing binding, the witness tier that detects it, and the repair primitive with its own denominator. The table explains why systems as different as a chat wrapper, a memory exporter, a KV scheduler, and a tool dispatcher belong to one analysis. In each case the upstream component was correct about its own record, and what was missing was a check by the consumer that the record still described the object it was about to use.

\subhead{Recovery must keep what is valid}
Rechecking at the consumer is necessary but not sufficient, because a consumer can make every check pass by discarding state. Clearing the ChatGLM2 cache meets exclusion and removes the history a long task depends on, and on the vLLM path eviction bounds admission while the continuation stays corrupted (Proposition~\ref{prop:conjunctive}). Exclusion, retention, and completion are therefore evaluated together as the recovery-adequacy criterion of Section~\ref{sec:contract-def}, and a repair is adequate only where all three were measured. Equation~\ref{eq:problem} covers the first two, which concern selection. Completion is the conditional progress obligation that the evaluation adds, since correct selection does not guarantee that a task succeeds. The authority side has the same shape with different predicates: a revision-bound dispatcher must refuse conflicting grants and still apply current and re-authorized ones (Table~\ref{tab:kappa-binding}).

\subhead{Evidence must follow the consumer and the predicate}
Deciding whether a repair is adequate requires evidence about the right request and the right predicate. Output similarity, request success, a hit count, and a component's success flag each misjudge some state predicate (Section~\ref{sec:tiering}). The BF16 control and the request-pairing experiment fail for different reasons. One violates Semantic Sufficiency and the other violates Identity, so a test that fixes arithmetic still fails if it inspects the wrong request (Proposition~\ref{prop:evidence}). A test of recovery reads the consuming request's own production range, allocation generation, execution position, and resource revision. Even then, a verifier that reads recorded metadata establishes conformance to those records and not the contents of device memory.

\subhead{Two instances of one principle}
State validity and authority freshness fail in the same way: a downstream operation treats an upstream decision as sufficient evidence and never re-establishes what it requires. What differs is the identity checked and the evidence that decides it (Table~\ref{tab:instances}). State predicates are decided at Tier~2, while dispatch effects need Tier-3 receipts. The predicates stay separate because the checks protect different consumers. Revision checks in a dispatcher do not protect a prefix index, and allocation generations in a cache do not protect a dispatcher.

\subhead{What a runtime should expose}
A consumer can make these checks only if the runtime gives it the identities to check. Table~\ref{tab:principles} turns these observations into interface requirements for a stateful agent runtime whose consumers must verify recovery rather than trust a completion flag. A runtime that reports only that a cancellation, rollback, or approval was processed leaves the consumer unable to tell which spans were fully produced, which allocation identities still name their object, and which resource versions still satisfy a grant. It should instead return the allocation epoch, completion receipt, resource revision, and rewind epoch that the consumer checks. Five requirements rest on a failure and a repair measured on a pinned system. Retirement and the separation of $t_{\mathrm{cancel}}$, $t_{\mathrm{settle}}$, and $t_{\mathrm{retire}}$ in the acknowledgment API follow from the lifecycle analysis of Section~\ref{sec:lifecycle} without a dedicated experiment, and we state them as design implications. These are requirements for a checkable recovery boundary, not a universal recovery monitor.

\subhead{Where the requirements apply}
The requirements assume that the runtime can expose an allocation epoch, a resource revision, and a log epoch. Where a provider hides its scheduler, Tier-2a witnesses are unavailable to the client, but rewind filtering and revision-bound dispatch still work in client-owned runtimes. What carries over to other runtimes is the question each consumer must answer, namely whether the state or authority it relies on is still valid at use, and not the specific predicates we measured.

\section{Related Work}
\label{sec:related}

\subhead{Recovery and rollback in stateful agents}
An earlier study identified rollback consistency for language agents: when an application keeps a session's KV cache across a logical abort, the model attends to a branch that the transcript no longer contains, which a same-token, different-cache audit separates from the visible text~\cite{abortedbut}. That study concerns logical history over a retained session handle and places content-addressed prefix caches such as vLLM's outside its channel, because they reuse only prefixes present in the request. Building on that observation, we examine recovery in asynchronous serving. There the requested prefix can match the retained history while its production, allocation identity, owner settlement, or execution position is invalid, so a logical-history check passes and the consumer still reads invalid state. We add consumer-bound contracts and an admissibility standard for the evidence that decides these conditions, authority freshness at tool dispatch, and a patch evaluation and factorial that test repairs against them.

Other work on agent recovery checks whether a restore is safe or builds mechanisms for it. Safe to Resume identifies failure modes of agent checkpoint and rollback, including stale external dependencies~\cite{saferesume}, and exact checking for execution edits decides whether a checkpoint, fork, restore, or merge is safe given the execution record~\cite{executionedits}. PGSR crops the KV cache to a clean checkpoint and replays a sanitized suffix~\cite{pgsr}, Leyline removes or replaces cached spans and corrects later positions~\cite{leyline}, and ChronoMem and Planarian version agent memory and state~\cite{chronomem,statepoints}. Reconstruction of this kind would satisfy our contracts. We test native recovery instead, and ask whether the consumers of an existing runtime hold valid state afterward.

\subhead{Memory, provenance, and authority freshness}
Revocation studies show that a retrieval guard can return a revoked record~\cite{revoked}, and other defenses bind records to their origin, attach typed provenance, or measure memory poisoning and authorization laundering~\cite{originbound,typedassertions,memsecbench,ealbench}. Planfence checks a stored plan's dependency frontier against current heads before a protected action~\cite{planfence}, and a memory-repair study gates on changes between snapshot and commit~\cite{evidencechanges}. Revision-bound dispatch applies the same freshness idea at an approval runtime's dispatcher. Those works govern stored records and plans. Our contracts also govern runtime representations such as prefix-index entries, block tables, and position counters, which matters because these representations can be invalid while every stored record is correct.

\subhead{KV caching and serving fault handling}
Serving security work studies content-driven leakage and poisoning: prefix timing side channels, prompt recovery from KV sharing, and poisoned reuse of cached reasoning~\cite{cachemecatch,promptpeek,r2ecap}, and providers document cross-request exposure and isolation settings~\cite{cloudflare,sgsecurity,vllmsecurity}. That work assumes the cached content itself is the attack vector. Our adversary supplies benign content and acts through lifecycle timing, which per-tenant salts leave open (Table~\ref{tab:mitigations}). Public trackers report several mechanisms we measure as functional defects: vLLM abort ordering, late reports after reuse, preempted owners, and the V2 recompute position~\cite{vllmrace,vllmfix,vllmreuse,vllmpreempt,vllmpreemptfix,vllmrewind}, aborted and intake-rejected SGLang prefill~\cite{sgretire,sgadmit,sgsourcehold}, and ADK export, Transformers caches, and a nanochat row guard~\cite{adkissue,hfphase24841,nanochat736}. Relative to them, we contribute the contract that unifies the cases and explains why PR 61769 satisfies 2 of 7 fault predicates, live reachability through a real connector to a second tenant's tool call, and the factorial showing that eviction and device rewind are jointly required.

\subhead{Complete mediation and concurrency control}
Complete mediation and the reference monitor require that every access, recovery included, be checked~\cite{saltzer1975,anderson1972}. TOCTOU analysis finds the hazard wherever a check and a use are separated~\cite{bishop1996}, and optimistic validation, lease generations, and grace periods bind a check to its object~\cite{occ,chubby,rcu}. These principles say that an access must be checked and that a check must be bound to the object it guards. They leave open what the check must establish when logical recovery, physical state, asynchronous execution, and downstream use come apart, as they do in the runtimes we study. We do not propose a new mechanism. We apply these established ones at boundaries where they were not in place (an aborted-load index entry, a recycled block identifier, a host-device counter pair, a rewound log, a queued grant), identify the consumer each must bind, and state the resulting conditions as contracts that admissible witnesses can decide. The experiments add what a restatement of the principles cannot: they show that these boundaries are left unchecked in pinned implementations, that a community patch closes some of the conditions and not others, and that two repairs are each required on the tested path.

\subhead{Test oracles and numerical nondeterminism}
Testing theory separates running a test from the soundness of its oracle~\cite{testoracle,testoracle2,difftest,difftest2,metamorphic}, and inference work attributes output nondeterminism to batch-dependent reduction order and builds batch-invariant kernels~\cite{batchinvariant,numdet,batchinvvllm}. The security consequence is that output equality is not an admissible witness for state recovery, with or without such kernels.

\section{Scope, Limits, and Ethics}
\label{sec:limits}

\subsection{What the evidence supports}
\label{sec:affects}

\subhead{Systems and versions}
The evaluation establishes reachability and causal structure on pinned stacks, not prevalence, and the patch comparison holds for the pinned trees and predicate set of Section~\ref{sec:state-obligations}. The serving models are small dense decoders (Qwen3-1.7B-Base, Phi-3.5-mini-instruct, ChatGLM2-6B in FP16); larger, mixture-of-experts, and FP8 or INT8 setups are untested.

\subhead{Fault model and evidence modes}
Faults are injected at fixed positions. The live approval evidence is one admitted episode, and the rest comes from controlled models. The 50 and 40 counts of Table~\ref{tab:tiering} are not a drift rate.

\subhead{Trace checking and instrumentation}
Lemma~\ref{prop:tier2a} tests reported metadata and informs about device state only when scheduler bookkeeping matches device memory. The $42/42$ agreement is conformance on recorded traces, and Tier-2b digests were not collected on the 100-run corpus.

\subhead{Generated calls and execution}
The two-tenant protocol establishes stale admission and wrong-target call generation, not an unauthorized write. The design requirements of Section~\ref{sec:principles} are inferences, marked in Table~\ref{tab:principles}.

\subhead{Scope of the theory}
Definition~\ref{def:ctv} fixes the form of a contract and not its predicates, and we give predicates only for the consumers studied. Propositions~\ref{prop:ack} and~\ref{prop:conjunctive} are existence results exhibited on pinned stacks, and Proposition~\ref{prop:evidence} covers the predicates we tested; none is a proof over arbitrary runtimes. The lifecycle phases of Section~\ref{sec:lifecycle} place predicates. Validity over a full asynchronous read and the draining of outstanding readers before reuse were not measured.

\subhead{When the contract is needed}
The need depends on whether state or authority can change between publication and use, not on tenancy (Section~\ref{sec:consumers}).

\subsection{Disclosure and artifact}

All experiments run in isolated local environments with synthetic tenants and benign canaries and touch no live third-party service. We plan to report the cross-tenant consequence, the incompleteness of PR 61769 under our fault predicates, and revision-blind dispatch, each with its repair, to the affected maintainers under coordinated disclosure; no notification has been completed at submission. The anonymized artifact, with pinned commits, frozen event logs, and the GPU-free verifier, is planned and not yet public; it holds no payloads beyond the benign canary protocol.

\section{Conclusion}
\label{sec:conclusion}

A cancellation, rollback, or approval is recorded by one component and relied on by another, and the record does not make the later use valid. The question this paper asks is how a stateful AI runtime can ensure that a downstream consumer uses only state and authority that remain valid when it consumes them. On pinned stacks, an interrupted load led a second tenant to generate a call naming the wrong tenant, and an approved write outlived the revision it was approved against. These are two forms of one failure: the consumer relied on an upstream decision and did not check the identity of the object it used. State validity and authority freshness differ in what must be checked, and they share the requirement that validity be established at the point of consumption. Consumer-bound contracts state what each consumer must establish. Admissible witnesses decide, from the consuming request's own metadata and artifacts, whether it holds. Boundary repairs make it hold without discarding valid state. The evidence is bounded to the pinned implementations and fault protocols we tested, and it shows necessity under those conditions and not a universal guarantee. Recovery interfaces of stateful AI runtimes should therefore let each consumer verify the conditions its use depends on.

\FloatBarrier
\begingroup

\endgroup

\appendix
\FloatBarrier
\FloatBarrier

\section{Numerical Protocol}
\label{app:numeric}

\subsection{Per-projection replay and the BF16 error envelope}
\label{app:numeric-protocol}

The fault-free divergence of Section~\ref{sec:oracles} needs a numerical
mechanism, not an assertion. Each arm captures live forward inputs and replays
every row in isolation and in batch across nine linear modules covering QKV,
attention output, and MLP projections in layers 0 and 27 alongside the
language-modeling head. Native kernel outputs match direct \code{F.linear} in all
18 captures except the two MLP down-projections of shape $2048\times6144$.
Substituting batch-invariant kernels on those two projections alone restores full
cold-reference agreement, and substituting every other linear projection leaves
the divergence intact. Repeating each captured row across 12 batch sizes from 1
to 2048 in fresh processes across three kernel implementations yields 288 checks.
Independent signed-integer accumulation of BF16 products yields exact dyadic
rationals for all 49 coordinates that show single-versus-batch variation,
including 26 nonzero FP64 discrepancies with maximum magnitude
$3.095\times10^{-15}$, far inside the $10^{-10}$ screening margin of the standard
floating-point error bound:
\begin{equation}
\begin{split}
|\widehat y_j-y_j^{(64)}| &\leq
\gamma_K\sum_{k=1}^{K}|x_k w_{jk}|\\
&\quad + \tfrac{1}{2}\,\mathrm{ulp}_{\mathrm{BF16}}(y_j^{(64)}) + 10^{-10},
\end{split}
\label{eq:envelope}
\end{equation}
where $\gamma_K=Ku/(1-Ku)$ and $u=2^{-24}$. Every observed discrepancy satisfies
Equation~\ref{eq:envelope} and reflects standard BF16 reduction-order sensitivity.

\subsection{Stale-output guard removal control}
\label{app:guard-control}

The guard control of Section~\ref{sec:state-obligations} is a single-line removal
ablation that replaces the drop-mode discard's \code{continue} with \code{pass};
a recursive diff across both source trees confirms the one-line delta.
Table~\ref{tab:guard-control} reports all 24 executions. The ``Armed'' column
reads the guard boolean from the live request struct, and the bottom rows verify
whether a synthetic failed-step token is delivered or committed under
\code{fail}.

\begin{table}[!htbp]
\centering
\caption{\textbf{The stale-output guard arms all $6/6$ fault cases, and removing its single line delivers and commits the stale token.} Stale counts discarded stale tokens; Armed records live request-object guard state, with $\checkmark$ armed and $\times$ unarmed.}
\label{tab:guard-control}
\scriptsize
\setlength{\tabcolsep}{4.2pt}
\renewcommand{\arraystretch}{0.98}
\begin{tabularx}{\columnwidth}{@{}X l r c r c@{}}
\toprule
& & \multicolumn{2}{c}{\textbf{Baseline}} & \multicolumn{2}{c}{\textbf{Guard Removed}}\\
\cmidrule(lr){3-4}\cmidrule(lr){5-6}
Trigger route & Policy & Stale & Armed & Stale & Armed\\
\midrule
None, live control & fail, rec. & 0 & $\times$ & 0 & $\times$\\
\code{reset\_prefix} & fail, rec. & 16 & $\checkmark$ & 16 & $\checkmark$\\
Preemption & fail, rec. & 16 & $\checkmark$ & 16 & $\checkmark$\\
\midrule
\emph{Token delivered} & fail & \multicolumn{2}{c}{$\times$} & \multicolumn{2}{c}{$\checkmark$}\\
\emph{Token committed} & fail & \multicolumn{2}{c}{$\times$} & \multicolumn{2}{c}{$\checkmark$}\\
\bottomrule
\end{tabularx}
\end{table}

\begin{table}[!htbp]
\centering
\caption{\textbf{Batch-invariant mode substitution preserves $100\%$ of Tier-2 state verdicts while generated tokens vary across modes.} Each frozen cell is re-run with \code{VLLM\_BATCH\_INVARIANT=1} and with default kernels. \emph{Probe}, \emph{Disp.}, and \emph{Suff.} are Tier-2 state witnesses.}
\label{app:rejudge-table}
\scriptsize
\setlength{\tabcolsep}{4.8pt}
\renewcommand{\arraystretch}{0.96}
\begin{tabularx}{\columnwidth}{@{}X c c c l@{}}
\toprule
& \multicolumn{3}{c}{\textbf{State verdict, all modes}} & \textbf{Tokens}\\
\cmidrule(lr){2-4}
Evaluated cell & Probe & Disp. & Suff. & Def./inv.\\
\midrule
Allocation-aware candidate & \checkmark & $\times$ & \checkmark & differ\\
No-device-rewind ablation & \checkmark & $\times$ & \checkmark & differ\\
Backported PR 61769 & $\times$ & $\times$ & $\times$ & differ\\
Unpatched upstream & $\times$ & $\times$ & $\times$ & differ\\
Unpatched, abort case & $\times$ & \checkmark & $\times$ & n/a\\
\midrule
Admission controls, 4 cells & \checkmark & \checkmark & n/a & ident.\\
\bottomrule
\end{tabularx}
\end{table}

\subsection{Mode-substitution re-judgment}
\label{app:rejudge}

Section~\ref{sec:rejudge} re-executes each output-flagged GPU cell under
\code{VLLM\_BATCH\_INVARIANT=1} with freshly generated within-mode cold
references, summarized in Table~\ref{app:rejudge-table}. All four admission
controls reproduce their invariant cold references token for token, so
deterministic GEMM kernels preserve connector transport and fault injection.
Under \code{recompute}, both allocation-aware rows pass \emph{Probe} and
\emph{Suff.} in all three modes, and \emph{Disp.} holds on the \code{fail} abort
row where the owner terminates in error. Unpatched upstream and backported PR
61769 emit the degenerate \code{[525]}$\times 8$ canary in every arithmetic
mode, which places the physical state contamination outside floating-point
reduction order.

\subsection{Comparator stability across reruns and hosts}
\label{app:comparator-stability}

Comparator stability is measured across every multi-run configuration of the
16-run packaged cohort. Six runs satisfy the state-corruption predicate of
Section~\ref{sec:false-negative} and ten are clean, and raw admitted-token counts
cannot separate them: consumer admission equals 256 in all 14 output-producing
runs, and probe admission equals 256 in 11 of the 16 runs.
Table~\ref{tab:comparator-stability} shows that the same-request Tier-3 primary
discriminator stays invariant across all six repeated configurations while the
cross-paired consumer continuation varies in three. On unpatched tree
\code{3b6dc718} under stale-KV recompute, four identical runs yield four distinct
consumer continuations, one of which matches the cold reference by chance, while
the probe's persisted artifact records \code{[525]}$\times 8$ in all $4/4$ runs.
Re-executing the 16-run protocol twice on a second H20 host with disjoint GPU
UUIDs reproduces the exact state partition, admitted-token counts, and verdicts
on $16/16$ runs for the primary model and for a second model family, whose
artifact bytes match on all $10/10$ clean runs and diverge on all $6/6$ corrupted
runs.

\subsection{Request-identity pairing ledger}
\label{app:pairing}

Table~\ref{tab:pairing-protocol} records all 16 packaged runs underlying
Section~\ref{sec:false-negative}. ``Adm.'' reports shared-prefix tokens admitted
to the later probe without computation. ``Corr.'' records whether aborted suffix
hashes stayed indexed, ``Prim.'' scores the probe's own persisted artifact
against the cold reference, and ``Sec.'' scores the original consumer's output as
``none'' on the two fail-closed runs that halt consumer output. Among the eight
clean runs that produce consumer output, two diverge from cold through batching
drift. Archive directory keys map \code{local\_c1} to the Allocation-Aware
Candidate, \code{public\_transplant} to Backported PR 61769, and \code{upstream}
to unpatched vLLM.

\begin{table}[!htbp]
\centering
\caption{\textbf{The per-run pairing ledger across all 16 packaged executions separates $6$ corrupted runs from $10$ clean runs.} Prim. pairs probe state with the probe's own persisted artifact; Sec. cross-pairs the original consumer output against the same cold reference.}
\label{tab:pairing-protocol}
\scriptsize
\setlength{\tabcolsep}{2.8pt}
\renewcommand{\arraystretch}{0.98}
\begin{tabularx}{\columnwidth}{@{}l X l l r c c c@{}}
\toprule
Arm & Tree & Cond. & Pol. & Adm. & Corr. & Prim. & Sec.\\
\midrule
Cache-only & Upstream & fail. & recomp. & 256 & none & match & match\\
Cache-only & Upstream & fail. & recomp. & 256 & none & match & match\\
Cache-only & Upstream & fail. & recomp. & 256 & none & match & match\\
Repaired & Alloc.-aware & fail. & recomp. & 64 & none & match & differ\\
Stale KV & Alloc.-aware & fail. & recomp. & 64 & none & match & differ\\
Stale KV & Alloc.-aware & fail. & recomp. & 64 & none & match & match\\
Stale KV & PR 61769 & fail. & recomp. & 256 & yes & differ & match\\
Stale KV & PR 61769 & fail. & recomp. & 256 & yes & differ & differ\\
Stale KV & Upstream & fail. & fail & 64 & none & match & none\\
Stale KV & Upstream & fail. & fail & 64 & none & match & none\\
Stale KV & Upstream & fail. & recomp. & 256 & yes & differ & differ\\
Stale KV & Upstream & fail. & recomp. & 256 & yes & differ & match\\
Stale KV & Upstream & fail. & recomp. & 256 & yes & differ & differ\\
Stale KV & Upstream & succ. & recomp. & 256 & none & match & match\\
Stale KV & Upstream & succ. & recomp. & 256 & none & match & match\\
Stale KV & Upstream & fail. & recomp. & 256 & yes & differ & differ\\
\bottomrule
\end{tabularx}

\end{table}

\begin{table}[!htbp]
\centering
\caption{\textbf{The same-request Tier-3 discriminator stays stable in all six repeated configurations while cross-paired output varies in three.} Grouped by arm, condition, failure policy, and commit prefix across 15 repeated runs.}
\label{tab:comparator-stability}
\scriptsize
\setlength{\tabcolsep}{3.5pt}
\renewcommand{\arraystretch}{0.95}
\begin{tabularx}{\columnwidth}{@{}l X r c c@{}}
\toprule
Arm, condition & Policy, tree hash & \textit{n} & Primary & Secondary\\
\midrule
cache\_off, failure & recompute, 3b6dc7 & 3 & stable & stable\\
stale\_kv, failure & fail, 3b6dc7 & 2 & stable & stable\\
stale\_kv, failure & recompute, 30592e & 2 & stable & stable\\
stale\_kv, failure & recompute, 3b6dc7, 62691c & 6 & stable & varies\\
stale\_kv, success & recompute, 3b6dc7 & 2 & stable & stable\\
\midrule
\multicolumn{5}{@{}l}{\scriptsize Total: 15 runs across 6 configurations; primary varies in 0, secondary in 3.}\\
\bottomrule
\end{tabularx}
\end{table}

\section{Oracle Audit Protocol}
\label{app:audit}

The oracle audit evaluates the 42-cell frozen GPU matrix, spanning seven source
arms and six fault conditions. Thirty-nine cells enter endpoint evaluation and
define 210 obligation slots: 28 five-obligation endpoint cells and 14 block-reuse
cells with four of five obligations applicable, for 196 applicable slots.
Re-running the standalone trace checker with all harness labels stripped infers
obligation structure from raw event streams alone and evaluates 199 entries,
because three reuse cells ended at a CUDA execution failure before emitting reuse
tags. Across the 89 obligations carrying direct harness booleans in
label-stripped mode and the 162 comparable obligations in full audit mode, label
removal changes zero verdicts, and raw event streams alone yield $89/89$
self-consistency. A cell passes when all applicable obligations hold, and an
obligation whose precondition never occurs in a trace is recorded as
obligation-level inconclusive. Table~\ref{tab:obligation-sets} summarizes audit
agreement and defines Obligation Sets A and B.

\begin{table}[!htbp]
\centering
\caption{\textbf{The standalone checker and blind cross-vendor parsers reproduce every endpoint verdict on the 42-cell matrix, with a $0/13$ false-witness rate and $3/149$ conservative false alarms.} Top: agreement against reference verdicts, where FW denotes false-witness probability over invalid obligations and FA the conservative false-alarm rate over $n$ comparable items. Bottom: the formal obligation sets derived from raw event streams.}
\label{tab:oracle-audit-summary}
\label{tab:obligation-sets}
\scriptsize
\setlength{\tabcolsep}{3.4pt}
\renewcommand{\arraystretch}{0.95}
\begin{tabularx}{\columnwidth}{@{}X r c c c@{}}
\toprule
Audit instrument and axis & \textit{n} & Agreed & FW & FA\\
\midrule
Standalone checker, obligations & 162 & 159, 98.1\% & 0/13, 0.0\% & 3/149, 2.0\%\\
Standalone checker, endpoints & 42 & 42, 100.0\% & 0/11, 0.0\% & 0/31, 0.0\%\\
Cross-vendor parsers, obligations & 158 & 158, 100.0\% & 0/13, 0.0\% & 0/145, 0.0\%\\
Cross-vendor parsers, endpoints & 35 & 35, 100.0\% & 0/8, 0.0\% & 0/27, 0.0\%\\
\bottomrule
\end{tabularx}
\par\medskip
\begin{tabularx}{\columnwidth}{@{}l X@{}}
\toprule
Set & Conjunctive trace obligations\\
\midrule
Set A: Endpoint & Fault reaches the engine; revoked key-value blocks are neither served nor left hashed; the valid prefix stays reusable; the original owner reaches its required terminal state; the later probe matches the cold reference.\\
\addlinespace[2pt]
Set B: Reuse & The coordinator processes the delayed report while sparing reallocated blocks; reused blocks advance allocation generation $g$; timestamps satisfy load $<$ abort $<$ reallocation $<$ report settlement; the new owner matches the cold reference; the original owner aborts cleanly.\\
\bottomrule
\end{tabularx}
\end{table}

\subhead{Zero dependence on harness labels and cross-host replay}
Every cell is adjudicated twice, once with full run metadata and once with the
injected condition, failure policy, and cell identifiers stripped, and the two
passes agree on all 42 cells. Executing the identical seven-tree cohort on a
second independent H20 host with disjoint GPU UUIDs reproduces all $42/42$ cell
verdicts.

\subhead{Cross-vendor blind specification parsers}
Two temperature-zero language models from independent vendor families receive
redacted raw event packets, reference token sequences, and probe records with all
harness booleans and checker outputs removed. Both return obligation verdicts
backed by explicit event sequence numbers, with $100\%$ inter-rater agreement.

\section{Coverage Registers}
\label{app:coverage}
\label{app:coverage-sec}

Table~\ref{tab:coverage} indexes evidence by structural \emph{coverage unit} and
separates planned \emph{conditions} from repeated \emph{executions}. Across every
studied stack the register holds 29 coverage units: seven mediation failures, 21
boundary controls or repairs, and one insufficient-witness path, of which 23
remain inside Google ADK's documented quiescence preconditions. A units column
reconciles those 29 stack by stack, so each of ChatGLM2, Google ADK, Cline,
vLLM, and SGLang carries a stated count and the class counts sum to the register.
The DSH approval runtime appears under its own denominator, since its eight live
episodes and seven archived replays form the authority-side cohort of
Section~\ref{sec:stale-writer} and are registered apart so episodes repeating one
operation contract do not inflate the count.

\begin{table*}[!htbp]
\centering
\caption{\textbf{The complete register of 29 coverage units separates seven mediation failures from 21 boundary controls and one insufficient-witness path, with every denominator kept unpooled.} Size reports distinct conditions unless marked as executions, histories, pairs, or trials. Units counts coverage units per row and sums to 29; the DSH approval runtime carries its own denominator and is not counted in that total.}
\label{tab:coverage}
\scriptsize
\setlength{\tabcolsep}{2.8pt}
\renewcommand{\arraystretch}{0.86}
\begin{tabularx}{0.988\textwidth}{p{0.255\textwidth} p{0.072\textwidth} c p{0.235\textwidth} X}
\toprule
Coverage unit and target stack & Size & Units & Denominator and outcome & Retained scope and methodological bound\\
\midrule
Branch deletion, ChatGLM2 & 9 exec. & 1 & Per-execution; Mismatch & Forward KV digests match the deleted branch in $9/9$\\
Naive cache clear, ChatGLM2 & 18 exec. & 1 & Per-execution; Mismatch & Incremental builder drops retained context in $6/6$ tasks\\
Streaming tail, ChatGLM2 & 1 source & 1 & Single-source trace; Correct & Full-history rebuild returns the complete cache\\
ADK parent view, native, filtered, rebuilt & 3 units & 3 & Per-unit; Mismatch, Repaired & Native projection mismatches, filtered and rebuilt views consistent\\
Active-event exporter and artifact, ADK & 9 hist. & 5 & 9 histories, 36 cond.; Mismatch, Repaired & Native $3/9$ against filtered and commit \code{3640e7d0c74b} $9/9$, artifact consistent\\
ADK late-completion set & 5 units & 5 & Per-unit; 1 Mismatch, 4 controls & Late native mismatch; complete-before-rewind, cancel, rebuild, fence controls\\
ADK durable SQLite set & 4 units & 4 & Per-unit; all correct & Durable policy; never-exposed and cancel-before-completion controls\\
Invalid load and PR 61769, vLLM & 96 trials & n/a & 18 native, 18 patched; Mismatch, Partial & Evidence for the three vLLM units below; abort-queue and direct-abort paths separated\\
vLLM cancellation, native and patched & 2 units & 2 & Per-unit; Mismatch, Repaired & Public patch closes the direct-abort owner path\\
Recompute path, vLLM & 6 exec. & 1 & Per-execution; Mismatch & Host-device position split after CPU-counter rewind\\
Rejected prefill retirement, SGLang & 4 runs & 1 & Paired schedules; Insufficient & Router cleanup bounded by the asynchronous timeout window\\
Cline checkpoint restore set & 5 units & 5 & Per-unit; all correct & Restore, repeat restore, late append, compaction, native effect\\
\midrule
\multicolumn{2}{@{}l}{\textbf{Register total}} & \textbf{29} & \textbf{7 Mismatch, 21 Correct, 1 Insufficient} & \textbf{Sums exactly; five stacks, no unit counted twice}\\
\midrule
Step-level witness checks & 10{,}112 & n/a & Forward steps audited; Pass & History, positional offset, and slot invariants\\
Unadmitted GPU setup cells & 5 cond. & n/a & Precondition $\op{Pre}$ unmet; Excluded & Preemption race, async transport drop, and DMA drain\\
\midrule
\multicolumn{5}{@{}p{0.988\textwidth}@{}}{\textbf{DSH approval runtime, separate denominator: 8 live episodes, 5 reaching dispatch evaluation, 3 stopping before $\op{Pre}$.}}\\
\multicolumn{5}{@{}p{0.988\textwidth}@{}}{\quad $1/1$ live admitted execution and $7/7$ archived provider replays; not pooled into the 29-unit register above.}\\
\bottomrule
\end{tabularx}
\end{table*}

\subhead{Hypothesis provenance and unpooled denominators}
\label{app:exposure}
Every coverage unit in Table~\ref{tab:coverage} carries source-informed
provenance, and the evidence modes it belongs to are described in
Section~\ref{sec:witness-cost}. Each experiment tests an explicit
mechanistic boundary hypothesis formulated from source or architectural
inspection before the cohort launches. Native GPU cohorts span 84 planned
conditions with 79 completions and 24 admitted fault cells. Agent cohorts span 42
SDK-projection episodes, 28 native runtime episodes, 163 approval requests across
24 conditions with 18 admitted, nine source-informed audit episodes, six
capacity-boundary tests, and eight subagent-manager conditions, all kept on
strictly unpooled denominators.

\begin{table}[!htbp]
\centering
\caption{\textbf{The endpoints re-stated after the oracle analysis of Section~\ref{sec:oracles} carry a named witness or a documented scope bound in all $12$ cases.} Each row binds an empirical endpoint to its witness class, upstream antecedent, and scope bound. Witness denotes direct Tier-2 or Tier-3 verification; Qualified marks a retained inequality with a documented batch-geometry scope.}
\label{tab:endpoints}
\label{app:affected}
\scriptsize
\setlength{\tabcolsep}{2.8pt}
\renewcommand{\arraystretch}{0.92}
\begin{tabularx}{\columnwidth}{@{}>{\raggedright\arraybackslash}p{0.33\columnwidth} >{\raggedright\arraybackslash}p{0.21\columnwidth} >{\raggedright\arraybackslash}X@{}}
\toprule
Endpoint or finding & Standing & Witness and scope bound\\
\midrule
Aborted prefix admitted at hit & Tier-2b / 3 Witness & 28-layer KV digests, 294 checks; $10/10$ wrong-target calls\\
Old report hits reused block & Tier-2a Witness & Block ID plus generation $g$ over the 42-case matrix\\
Owner reaches error status & Tier-2a Witness & Native \code{FINISHED\_ERROR} and token history\\
GPU position stale post-rewind & Tier-2a Witness & Host tokens at 128 against GPU position at 288\\
Recompute differs from cold & Tier-2 Witness & State verdict $100\%$ invariant across modes\\
New-owner output matches cold & Tier-2 Witness & Probe contamination mode-invariant; new owner clean\\
No-rewind arm differs from cold & Tier-2 Witness & Host\kern0.35pt-\kern0.35pt device split persists in invariant mode\\
Unprimed control differs & Qualified & Token inequality retained; batch confound noted\\
Stale-output guard drops token & Removal control & \code{continue} $\rightarrow$ \code{pass} commits the token in $6/6$ armed cases\\
\midrule
Exporter reads raw events & ADK 7570 & Active-view filter and \code{3640e7d0c74b}, 36 cond.\\
Stale prefix, reuse, preemption & vLLM 61768, 61905, 61912 & 96 GPU trials and 42 CPU cases against \code{b8c04dd3}, \code{8ee68eb1}\\
Output oracle unsound for state & Counterexample & 18-run factorial and 288-check BF16 MLP replay\\
\bottomrule
\end{tabularx}
\end{table}

\begin{table}[!htbp]
\centering
\caption{\textbf{Upstream PR 61769 closes running-owner invalidation and leaves \code{PREEMPTED} owners unresolved, so allocation-aware eviction with preempted settlement is the remedy that satisfies the owner obligation.} Hit / Exp. compares prefix tokens admitted to a later request against the 48 valid tokens preceding the failed block; Owner reports whether the preempted owner reaches terminal settlement.}
\label{tab:repair-tree}
\scriptsize
\setlength{\tabcolsep}{4.0pt}
\renewcommand{\arraystretch}{0.88}
\begin{tabularx}{\columnwidth}{@{}X c c c c@{}}
\toprule
& \multicolumn{2}{c}{\textbf{Hit / Exp.}} & \multicolumn{2}{c}{\textbf{Owner}}\\
\cmidrule(lr){2-3}\cmidrule(lr){4-5}
Evaluated source tree & Fail & Recomp. & Fail & Recomp.\\
\midrule
Unpatched upstream & 112/48 & 112/48 & bad & n/a\\
Backported PR 61769 & 48/48 & 48/48 & bad & n/a\\
Allocation-aware eviction & 48/48 & 48/48 & bad & n/a\\
\quad $+$ preempted settlement & 48/48 & 48/48 & ok & n/a\\
\bottomrule
\end{tabularx}
\par\smallskip
{\scriptsize Hit / Exp. is measured on the direct-abort case. On the preempted path the CPU matrix shows backported PR 61769 readmitting 112 tokens against 48 expected. ``Bad'' marks an owner left in \code{PREEMPTED} under \code{fail}, because legacy failure mapping scans only the running and remote-waiting queues. The owner-terminal predicate is not scored under \code{recompute}, where the owner is rewound and requeued by design, so a non-terminal owner there is expected behavior and not a violation (n/a). Both upstream commits were closed without merging (PR 61769 on November 12, 2026).}
\end{table}

\section{Native GPU and Agent Cohort Protocols}
\label{app:gpu}

\begin{figure*}[!htbp]
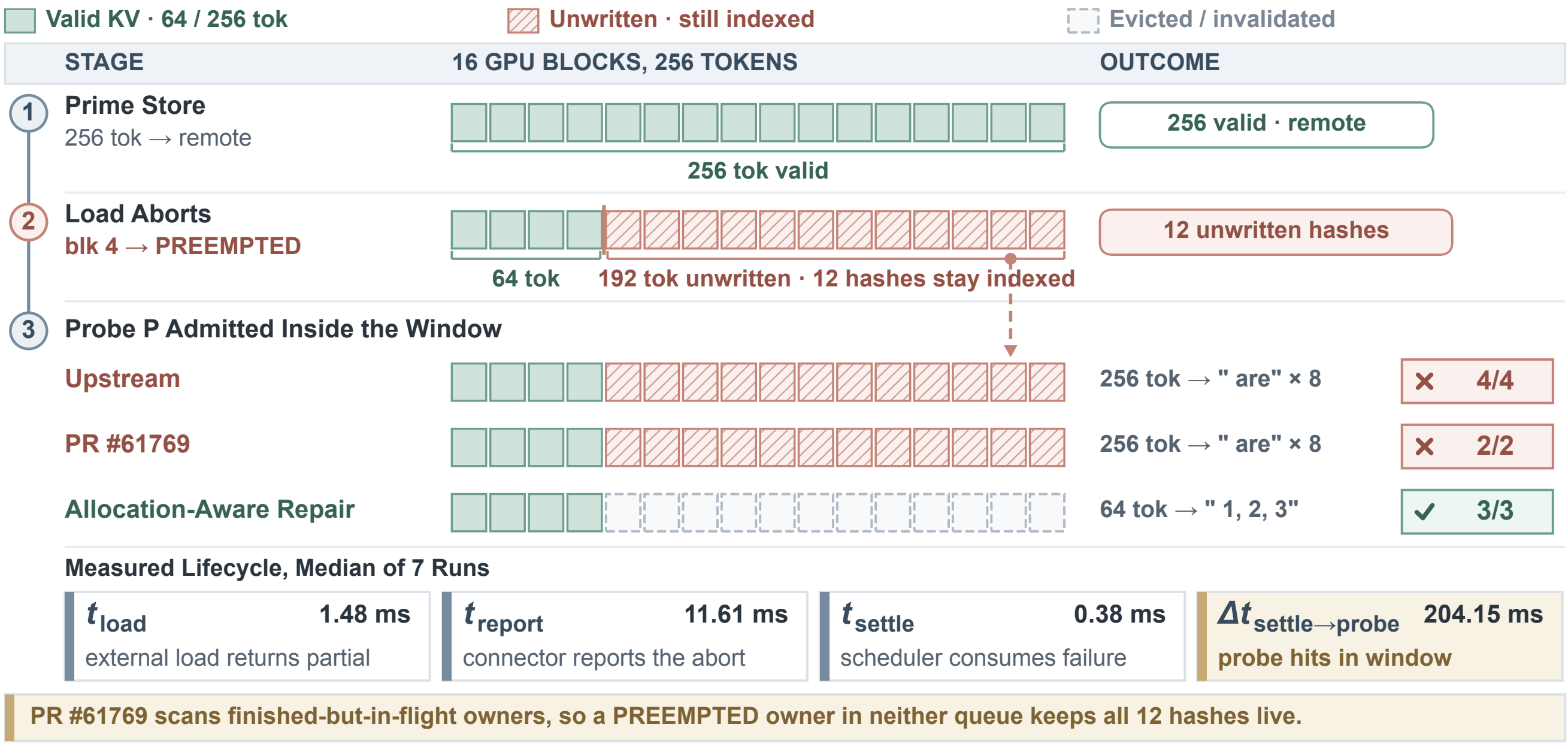

\centering
\PaperFigureFive{\textwidth}
\caption{\textbf{A failed external load leaves twelve unwritten block hashes indexed for later requests, and allocation-aware eviction plus device rewind together stop the contamination.} Consumer~C aborts across blocks 4 through 15 and enters \code{PREEMPTED}, so 12 unwritten suffix hashes stay live in the prefix index; probe~P is then admitted all 256 stale tokens on unpatched upstream and backported PR 61769, while allocation-aware repair restricts admission to the 64 valid tokens. Medians across 7 runs record the partial external load at 1.48~ms, the connector abort report at 11.61~ms, scheduler settlement at 0.38~ms, and the interval from failure consumption to probe output at 204.15~ms, labeled $\Delta t_{\mathrm{settle}\to\mathrm{probe}}$, over which the owner stays unretired and the stale hashes stay indexed.}
\label{fig:block-lifecycle}
\end{figure*}

\subhead{Eager-mode pointer-registration shim}
Figure~\ref{fig:block-lifecycle} and Table~\ref{tab:repair-tree} run vLLM under
\code{enforce\_eager} with a single-call connector pointer-registration hook
before each store, so LMCache V3 registers the physical tensor addresses that
default eager execution skips outside CUDA-graph capture.

\subsection{Agent cohorts and deferred-authority preconditions}
\label{app:agents-sec}

Agent-side cohorts test whether historical context or superseded permissions leak
across session boundaries. The eight-episode live-provider matrix of
Section~\ref{sec:stale-writer} crosses \code{allowed-once}, \code{canceled},
\code{rejected}, and \code{serialized} across two model families. Five episodes
reach dispatch evaluation and three fail precondition $\op{Pre}$: one provider
family emitted no pending write draft in two runs, and one \code{serialized} run
exceeded the setup timeout before draft creation. The planned artifact
(Section~\ref{sec:limits}) keeps the original gateway routing model identifiers so that
each episode can be replayed against the endpoint that served it.

\subsection{Root-cause sufficiency bridge}
\label{app:bridge}

The repaired-arm cohorts of Table~\ref{tab:sufficiency-bridge} are the six
matched-task conditions of Section~\ref{sec:consumers} (ChatGLM2), nine histories
(ADK), the ten $E\wedge D$ runs of Table~\ref{tab:kv-policy} and the two triggered
generation cases of the 42-case matrix (vLLM), the two patched-tree runs of four
reject-trigger runs (SGLang), and the 160 conflicting instances of
Table~\ref{tab:kappa-binding} (approval runtime).

\begin{table}[!htbp]
\centering
\caption{\textbf{Six cross-boundary failures collapse onto three missing bindings, and each repair is scored on the denominator of its own cohort.} Each row joins a consumer boundary to the violated binding, the witness tier from Table~\ref{tab:tiering} that decides it, the observed failure, and the repair primitive. The last column gives the repaired-arm outcome with its own unit and denominator; the units differ across rows and are not pooled. ADK cites Tier~3 because its evidence is the re-read export artifact, not attended-KV bytes.}
\label{tab:sufficiency-bridge}
\scriptsize
\setlength{\tabcolsep}{3.5pt}
\renewcommand{\arraystretch}{0.92}
\begin{tabularx}{\textwidth}{@{}l l >{\raggedright\arraybackslash}X c@{}}
\toprule
System & Boundary & Root-cause class and missing binding & Tier\\
\midrule
ChatGLM2 & hist.\ $\to$ KV & Use-after-revoke: epoch unbound to pass & 2a/2b\\
ADK & log $\to$ view & Use-after-revoke: rewind unbound to export & 3\\
vLLM & store $\to$ blk & Use-after-revoke: abort unbound to hashes & 2a\\
vLLM & rep.\ $\to$ blk & Alloc.\ binding: report omits generation $g$ & 2a\\
SGLang & sched.\ $\to$ wkr & Use-after-revoke: reject unbound to send & 2a\\
Approval & grant $\to$ res. & Revision TOCTOU: grant omits revision $\kappa$ & 3\\
\midrule
\multicolumn{4}{@{}l}{\emph{Structural repair primitives and their repaired arms.}}\\
\toprule
System & Observation & Structural repair primitive & Repaired arm\\
\midrule
ChatGLM2 & $9/9$ read deleted KV & Reconstruct $C_{:r}$ and $C_{r:}$ & $6/6$ cond.\\
ADK & $3/9$ against $9/9$, 36 cond. & Filter export by rewind epoch & $9/9$ hist.\\
vLLM & 96 trials; $6/6$ stale & Evict suffix and rewind GPU position & $10/10$ runs\\
vLLM & 42-case reuse matrix & Bind report to generation $g$ & $2/2$ cases\\
SGLang & 2/2 base runs forward a token & Gate KV send on acceptance & $2/2$ runs\\
Approval & $1/1$ live, $1{,}502$ race & Enforce $\kappa_{\mathrm{grant}} = \kappa_{\mathrm{now}}$ & $160/160$ inst.\\
\bottomrule
\end{tabularx}
\end{table}

\FloatBarrier


\begin{thebibliography}{99}
\setlength{\itemsep}{1pt}
\setlength{\parskip}{0pt}
\raggedbottom
\bibitem{greshake} K. Greshake et al., ``Indirect prompt injection in real-world LLM applications,'' arXiv:2302.12173, 2023.
\bibitem{agentdojo} E. Debenedetti et al., ``AgentDojo: A dynamic environment for prompt-injection attacks,'' arXiv:2406.13352, 2024.
\bibitem{pagedattention} W. Kwon et al., ``Efficient memory management for LLM serving with PagedAttention,'' arXiv:2309.06180, 2023.
\bibitem{sglang} L. Zheng et al., ``SGLang: Efficient execution of structured LM programs,'' arXiv:2312.07104v2, 2024.
\bibitem{leyline} B. Ma et al., ``Leyline: KV cache directives for agentic inference,'' arXiv:2606.01065v1, 2026.
\bibitem{pgsr} C. Yao et al., ``Forgetting without restarting: Execution-state unlearning for stateful LLM agents,'' arXiv:2609.04875v1, 2026.
\bibitem{revoked} Y. T. Shen et al., ``Revoked but still authoritative: Revocation enforcement in agent-memory systems,'' arXiv:2609.08258v1, 2026.
\bibitem{adkissue} ADK-Python maintainers, ``Rewound turns still stored in long-term memory,'' \code{google/adk-python} issue 7570, 2026.
\bibitem{space} mikeee, ``chatglm2-6b-4bit: app.py,'' Hugging Face Spaces; snapshot in the artifact.
\bibitem{cline} Cline contributors, ``Cline Core SDK: local runtime, checkpoint restoration,'' v0.2.0; snapshot in the artifact.
\bibitem{ohlocal} OpenHands contributors, ``LocalConversation: fork, navigate, confirmation APIs,'' SDK 1.58.2; archived with the artifact.
\bibitem{cloudflare} Cloudflare, ``Cross-request KV cache exposure in GLM-5.3 Workers AI and our remediation,'' 2026.
\bibitem{vllmrace} A.~Deb\kern0.25pt, vLLM issue 61768, \code{vllm-project/vllm}, 2026.
\bibitem{vllmfix} vLLM contributors, PR 61769, \code{vllm-project/vllm}; pinned patch in the artifact.
\bibitem{r2ecap} L. Zhu et al., ``Stealthy poisoning attacks via reused KV cache of reasoning,'' arXiv:2610.22873v1, 2026.
\bibitem{sgretire} SGLang contributors, ``Retire aborted disaggregated prefill results,'' PR 36988, \code{sgl-project/sglang}.
\bibitem{sgadmit} SGLang contributors, ``Do not admit intake-rejected requests to a PD handoff,'' PR 38935, \code{sgl-project/sglang}.
\bibitem{sgsecurity} SGLang Team, ``SGLang security update: Improving request isolation,'' 2026.
\bibitem{dsh} deepseek-ai, ``dsh-superagent research prototype,'' \code{deepseek-ai/dsh-superagent-prototype}, commit 6337fc823ce20251c8de8085948165c81aaa701e, Oct. 31, 2026.
\bibitem{dshruntime} deepseek-ai, ``DSH user approval and agent runtime,'' \code{deepseek-ai/dsh-superagent-prototype} v0.2.1-alpha.2; archived with lockfile.
\bibitem{storedinjection} Y. Xie et al., ``Rethinking agent security through cross-session stored prompt injection,'' arXiv:2606.04425v2, 2026.
\bibitem{memsecbench} X. Chen et al., ``MemSecBench: Tracking agent memory poisoning from persistence to consequence and repair,'' arXiv:2607.27080v1, 2026.
\bibitem{originbound} Y. Louck, ``Securing LLM-agent long-term memory against poisoning: Origin-bound authority,'' arXiv:2606.24322v1, 2026.
\bibitem{ealbench} T. Cerruti et al., ``Agent memory is a surface for endogenous authorization laundering,'' arXiv:2609.01836v2, 2026.
\bibitem{typedassertions} J. He and D. Yu, ``Stored is not supported: Typed provenance and assertion guardrails for persistent AI agents,'' arXiv:2609.02127v1, 2026.
\bibitem{planfence} E. Chen et al., ``Fresh memory, stale plans: Derivation currency for distributed LLM-agent memory,'' arXiv:2609.03340v2, 2026.
\bibitem{evidencechanges} W. Chu, ``When evidence changes: Evaluating memory repair and re-reading in language-model agents,'' arXiv:2610.03902v1, 2026.
\bibitem{saltzer1975} J. H. Saltzer and M. D. Schroeder, ``The protection of information in computer systems,'' \emph{Proc. IEEE}, vol. 63, no. 9, pp. 1278--1308, 1975.
\bibitem{anderson1972} J. P. Anderson, ``Computer security technology planning study,'' ESD-TR-73-51, Vol. II, Electronic Systems Division, AFSC, Hanscom Field, 1972.
\bibitem{bishop1996} M. Bishop and M. Dilger, ``Checking for race conditions in file accesses,'' \emph{Computing Systems}, vol. 9, no. 2, pp. 131--152, 1996.
\bibitem{occ} H. T. Kung and J. T. Robinson, ``On optimistic methods for concurrency control,'' ACM Trans. Database Syst., vol. 6, no. 2, pp. 213--226, 1981.
\bibitem{chubby} M. Burrows, ``The Chubby lock service for loosely-coupled distributed systems,'' in OSDI, 2006.
\bibitem{vllmreuse} A.~Deb\kern0.25pt, vLLM issue 61905, \code{vllm-project/vllm}, 2026.
\bibitem{vllmpreempt} guanxingithub, ``Load-failure report arriving after the owner was preempted,'' issue 61912, \code{vllm-project/vllm}, Oct. 2026.
\bibitem{vllmpreemptfix} vLLM contributors, ``Handle load failures after preemption,'' PR 62287, \code{vllm-project/vllm}; closed unmerged Nov. 13, 2026.
\bibitem{vllmrewind} ayoub-ibm, ``\code{kv\_load\_failure\_policy="recompute"} gives wrong output when a KV connector rejects a synchronous load (V2 runner),'' issue 49250, \code{vllm-project/vllm}, Jul. 2026.
\bibitem{sgsourcehold} SGLang contributors, ``Defer aborted prefill release until source reads drain,'' PR 47250, \code{sgl-project/sglang}; open as of Nov. 2, 2026.
\bibitem{hfphase24841} Hugging Face Transformers contributors, ``Caching prompt hidden states across \code{generate} calls,'' issue 24841.
\bibitem{nanochat736} nanochat contributors, ``Skip the row completed in the inference sample loop,'' PR 736, \code{karpathy/nanochat}.
\bibitem{rcu} Linux kernel contributors, ``RCU concepts,'' Linux v5.16 documentation.
\bibitem{batchinvariant} H. He, ``Defeating nondeterminism in LLM inference: True batch-invariant operations,'' Thinking Machines Lab, 2025.
\bibitem{numdet} J. Yuan et al., ``Understanding and mitigating numerical sources of nondeterminism in LLM inference,'' arXiv:2506.09501, 2025.
\bibitem{batchinvvllm} vLLM contributors, ``Batch-invariant kernels for vLLM,'' under \code{VLLM\_BATCH\_INVARIANT}.
\bibitem{metamorphic} M. Black et al., ``Metamorphic testing for generative AI,'' in \emph{Software Quality}, 2023, pp. 181--201.
\bibitem{difftest} M. Kiezun et al., ``Differential software testing,'' in \emph{Software Engineering: Methods and Analysis}, 2015, pp. 235--248.
\bibitem{testoracle} J. B. Goodenough and S. L. Gerhart, ``Toward a theory of test data selection,'' \emph{IEEE Trans. Softw. Eng.}, vol. SE-1, no. 2, pp. 156--173, 1975.
\bibitem{testoracle2} E. T. Barr, M. Harman, P. McMinn, M. Shahbaz, and S. Yoo, ``The oracle problem in software testing: A survey,'' \emph{IEEE Trans. Softw. Eng.}, vol. 41, no. 5, pp. 507--525, 2015.
\bibitem{difftest2} R. A. McKeeth, ``Testing differentially-based applications,'' in \emph{Proc. Int. Conf. Software Reliability and Quality}, 1997.
\bibitem{vllmsecurity} vLLM, ``Security: Multi-tenant deployments and prefix-cache timing mitigation,'' documentation.
\bibitem{cachemecatch} X. Wu et al., ``Cache me, catch you: Cache related security threats in LLM serving frameworks,'' in NDSS, 2026.
\bibitem{promptpeek} X. Zhang et al., ``I know what you asked: Prompt leakage via KV-cache sharing in multi-tenant LLM serving,'' in NDSS, 2025.
\bibitem{saferesume} G. Wu et al., ``Safe to resume? Breaking execution continuity of agent execution via rollback,'' arXiv:2608.29381v1, 2026.
\bibitem{chronomem} Y. Su et al., ``ChronoMem: Version control and semantic rollback for LLM agent memory,'' arXiv:2607.27773v1, 2026.
\bibitem{statepoints} J. Guo et al., ``Planarian: Managing agent state with statepoints,'' arXiv:2609.35366v1, 2026.
\bibitem{abortedbut} G. Zhang and H. Yang, ``Aborted but not forgotten: KV-cache retention breaks rollback consistency in language agents,'' arXiv:2608.15939v1, 2026.
\bibitem{executionedits} Y. Zheng et al., ``When can agents safely checkpoint, fork, restore, and merge? Exact checking for execution edits,'' arXiv:2608.22928v1, 2026.
\end{thebibliography}
\end{document}